\documentclass[11pt]{article}
\usepackage{jheppubmod}
\pdfoutput=1
\usepackage{psfrag}
 \usepackage{subfig}
\usepackage{comment}
\usepackage{array}
\usepackage{amssymb}
\usepackage{amsmath}
\usepackage{amsthm}
\usepackage{graphicx}
\usepackage{float}

\title{Dirac Born-Infeld inflation under constant roll conditions}
\author[a]{Sayantani Lahiri}

\affiliation[a]{ZARM, University of Bremen, \\
	Am Falltrum,
	28359 Bremen, Germany}
\emailAdd{sayantani.lahiri@zarm.uni-bremen.de}

\bigskip

\abstract{ In the present paper, we aim to generalize the constant-roll inflation in the non-canonical set-up, particularly in the context of scalar Dirac-Born-Infeld (DBI) theory which has emerged to be prospective candidate for describing inflation in the realm of string theory and in general permits time dependent speed of sound. Assuming the simplest case, when the speed of sound of scalar fluctuations is constant but different from unity, we obtain exact inflationary solutions and corresponding solutions of warp factors under constant-roll conditions. In this scenario, the inflationary solution is found to be an attractor. The power spectra of curvature fluctuations is analysed in the DBI regime and finally an estimation for predicting the observational bound on the DBI constant roll parameter is performed by using the the values of scalar spectral index and speed of sound provided by Planck and WMAP data.}

\begin{document}
	\maketitle

	\section{Introduction}	
	The standard big bang model which forms a cornerstone in cosmology is however plagued with several discrepancies like flatness problem, horizon problem, explanation for non-existence of magnetic monopoles, inability to explain temperature fluctuations in CMB, all of which remain unanswered in the realm of hot big bang model. 
	The inflationary paradigm \cite{Guth} that proposes an accelerated expansion of the early universe in a very short time interval successfully addresses all these issues. Moreover, the primordial scalar perturbations generated during inflation provides an explanation to the origin of the structure formation of the current universe \cite{Hawking} in addition to confirmation of homogeneity and isotropy of the universe on large scales \cite{Planck1,Planck2}. Well-supported by current observations \cite{Planck1, WMAP}, inflation forms an indispensable part for the description of the early universe together with the standard cosmological big bang model. \\
The simplest model of inflation involves a single scalar field also known as the inflaton field endowed with an inflaton potential and a canonical kinetic term (see reviews, \cite{Liddle, Baumann}) where the quasi de-Sitter expansion of the universe driven by the inflaton field generally relies on slow-roll approximation which is fulfilled if the kinetic energy of the inflaton field is negligibly smaller than the potential energy and the inflaton in this case slowly rolls down the nearly flat potential. Nevertheless, multi-field inflation models with more than one scalar field aiding the accelerated expansion of the universe have been widely studied \cite{Langlois1, Wands} in the past years. \\
On the other hand, inflationary scenarios with non-canonical kinetic term was first studied by Mukhanov and Garriga \cite{Mukhanov1} known as k-inflation models which are the non-canonical generalization of standard canonical inflation models. In contrast to canonical inflation models, where scalar fluctuations travel at the speed of light, the characteristic feature of the k-inflation models is the these fluctuations propagate with the speed of sound hence offer distinct perturbation spectra signatures as suggested by Mukhanov \cite{Garriga} in addition to other features like enhanced non-Gaussianity \cite{Planck1,Planck2} in primordial perturbations. Different variants of k-inflation models, models of non-canonical inflation and the corresponding observational realizations have been extensively studied \cite{Unni,Hu,Kinney1, Martin}. \\ 
Generally, inflation is believed to occur at energy scales when quantum corrections of gravity become important and to date the superstring theory provides the only consistent formulation of quantization of gravity involving extra spatial dimensions \cite{Green}. 
This Motivated by these developments, several models of inflation have been constructed in the framework of String theory aiming to provide natural realization and explanation of inflation. In this direction, one of the approaches involve identification of the inflaton field as the open string mode for example models of D-brane inflation \cite{Silverstein, Tong}, KKLMMT scenario \cite{KKLMMT}, induced brane inflation \cite{Dvali}. In particular, the D-brane inflation occurs with non-standard kinetic term where accelerated expansion of the universe can be visualized as a consequence of the relativistic motion of a $D3$ brane in the higher dimensional warped spacetime or the throat such that the low energy effective four dimensional theory on the D3-brane is described by Dirac Born-Infeld action (DBI). 
Extensive studies of inflationary scenarios associated to single field and multi field D-brane inflation, their implications on the power spectra, non-gaussianity have been carried out in recent years \cite{Kinney2, Spalinski1, Tye, Kob, Bruck, Pajer}.\\
In the recent work by Starobinsky \cite{Starobinsky1}, it has been shown that universe undergoes a phase of accelerated expansion under the influence of an inflaton field with a canonical kinetic term in absence of slow roll approximations. Although such a scenario has been previously proposed as ultra slow-roll inflation with exactly flat potential, the novel feature of the constant roll inflation is that exact inflationary solutions are obtained beyond slow-roll regime with a non-flat potential. 
These inflationary solutions in a certain parameter region have been shown to be an attractor with feasible observational realizations \cite{Starobinsky1, Starobinsky2, Waj}.
 The constant-roll inflation model has been generalized and extended for different theories \cite{Nozari,Mohammadi,Karam,Ito}. \\
In the present work, we aim to generalize the constant roll inflation model with non-canonical kinetic term and therefore we consider specifically the DBI term.
 Although the DBI action arises from the solution of the superstring theory\cite{Tong}, from the point of the view of studying constant-roll inflation in presence of non-standard kinetic term, we treat the DBI action as a phenomenological model.
From the phenomenological of view, the scalar DBI model belongs to the general class of k-inflation models proposed by \cite{Garriga} and hence shares several common characteristics namely time-dependent speed of sound, distinct power spactra of the primordial perturbations and prediction of enhanced non-Gaussianity \cite{Silverstein}. The DBI inflation with constant speed of sound \cite{Spalinski-2} as well as with a variable speed of sound has been extensively studied both in the context of slow-roll conditions and beyond \cite{Silverstein, Tong, Peris}.\\
In this work, we shall consider the simplest case when $\gamma$ which is inversely proportional to the speed of sound $c_s$, is constant, thus implying constant speed of scalar perturbations. 
Such a consideration relatively simplifies the equations of motion for determining the analytical solutions and at the same time facilitates the study of constant-roll inflation model and its cosmological implications with the non-standard kinetic term in the backdrop of DBI scenario.
The obtained exact inflationary solutions with constant  $\gamma$ but $\gamma \neq 1$ and the associated power spectra thereby help us explore the deviation from canonical formalism carried out in \cite{Starobinsky1, Starobinsky2}.\\
The present paper is organized as follows. In section 2 and section 3, we respectively recapitulate preliminaries of constant roll inflation for the canonical case and inflation with a single scalar field in the DBI scenario. 
In section 4, we  describe constant roll inflation in non-canonical set-up with the DBI action. Taking constant $\gamma$ where $\gamma \neq 1$, we have obtained exact inflationary solutions using the Hamilton-Jacobi formalism. However the obtained inflationary solutions parametrized by $\gamma$ and constant roll parameter defined in DBI scenario shows an attractor behaviour in a certain parameter region. We also analysed the corresponding power spectra of the scalar perturbations and estimated the value of constant roll parameter in this case. Finally, in section 5, we discuss our results. 
	
	\section{Canonical constant-roll inflation}
	The constant roll inflation model \cite{Starobinsky1} proposes that accelerated expansion of the universe can be sustained without employing slow roll approximations and without requiring the inflaton potential to be exactly flat \cite{ultra-roll}. The novel feature of the model is that inflationary solutions are obtained by solving equations of motion exactly. 
	These obtained solutions are characterized by a constant roll parameter whose value can be estimated using present observational bounds \cite{Starobinsky2}.
	  The constant roll inflation model described by a minimally coupled scalar field $\phi$ is given by the following action,
	\begin{equation}
	S=\int d^4 x \sqrt{-g}\left[\displaystyle \frac{R}{2 \kappa^2}-\displaystyle \frac{1}{2} g^{\mu \nu} \partial_{\mu} \phi \partial_{\nu} \phi-V(\phi)\right]
	\end{equation}
	where $V(\phi)$ is the corresponding inflaton potential and $\kappa^2$ is related to the four dimensional Newton's constant. 
	During inflationary phase, the universe is assumed to be isotropic, homogeneous and spatially flat spacetime described by the spatially flat FRW metric as follows,
	\begin{equation}
	ds^2= -dt^2+a^2(t)\delta_{ij}dx^{i}dx^{j}  \label{metric-1}
	\end{equation}
	The equations of motion are given by,
	\begin{eqnarray}
	H^2&=&\displaystyle \frac{\kappa^2}{3}\left[\frac{\dot{\phi}^2}{2}+V(\phi)\right] \\
	\dot{H} &= -&\frac{\kappa^2}{2} \dot{\phi}^2  \label{H-dot}\\
	\ddot{\phi}+3 H \dot{\phi}+V'(\phi)&=&0  \label{inflaton-canonical}
	\end{eqnarray}
	where dot denotes derivative with respect to $t$ and $'$ denotes derivative with respect to $\phi$.
	 When slow-roll conditions are not employed, the equation of motion of the inflaton field expressed in terms of Hubble flow parameter becomes,
	\begin{equation}
	\eta_{H}=-\displaystyle \frac{\ddot{\phi}}{H\dot{\phi}}=(3+\alpha)  \label{second}
	\end{equation}
	where
	\begin{equation}
	\alpha=\frac{V'(\phi)}{H \dot{\phi}}  \label{alpha-can}
	\end{equation}
	where the parameter $\alpha$ is called the constant roll parameter that denotes deviation from the flat potential and thus addresses the situation when $\eta \sim {\cal O}(1)$. Then $\alpha=0$ corresponds to the ultra slow roll case whereas the slow-roll inflation models can be retrieved for $\alpha\simeq-3$ implying  $\ddot{\phi} \rightarrow0$ and hence $\eta \rightarrow0$.
	The dynamics of the inflation field in absence of slow-roll conditions is studied using Hamilton-Jacobi formalism where the scalar field $\phi$ is taken as the time variable as a result the Hubble's function becomes a function of $\phi$. 
	Such a consideration in turn suggests that temporal co-ordinate is a single-valued function of $\phi$ i,e. $t=t(\phi)$ and $\dot{\phi}\neq 0$. Since $H=H(\phi)$, (\ref{H-dot}) can be readily expressed as,
	\begin{equation}
	H'(\phi)=-\displaystyle \frac{\kappa^2}{2}\dot{\phi}
	\end{equation}
	and together with (\ref{inflaton-canonical}) one obtains the differential equation of the Hubble parameter as,
	\begin{equation}
	H''(\phi)-\displaystyle \frac{\kappa^2}{2}(3+\alpha)H(\phi)=0  \label{Hubb-can}
	\end{equation}
	which is a second order differential equation and the solution of $H(\phi)$ is given by,
	\begin{equation}
	H(\phi)=C_1 e^{\sqrt{\frac{3+\alpha}{2}}\kappa \phi}+C_2 e^{-\sqrt{\frac{3+\alpha}{2}}\kappa \phi}  \label{solution}
	\end{equation}
where $C_1$ and $C_2$ are integration constants.
Then (\ref{solution}) can be analysed for two different parameter spaces namely $\alpha>3$ and $\alpha<-3$ for determining all allowed inflationary solutions subject to constant-roll conditions \cite{Starobinsky1, Starobinsky2}.
	\section{Inflation in DBI scalar theory}
	In this section we briefly describe inflation in the context of scalar DBI theory where the action on a D3 brane consists of non-canonical kinetic term. 
	A D3 brane specified by DBI action and moving in the six dimensional warped throat such that the line element of the ten dimensional spacetime corresponding to the warped flux compactification of the type IIB string theory is given by \cite{Peris} as,
	\begin{equation}
	ds^2=h^{-\frac{1}{2}}g_{\mu \nu}dx^{\mu} dx^{\nu}+h^{\frac{1}{2}}(dr^2+r^2ds^2_{5})  \label{warped-metric}
	\end{equation}
	where $h(r)$ is the warp factor and r is the radial co-ordinate. Here $\mu=0,1,2,3$ describe co-ordinates on the four dimensional world and five dimensional $ X_5$ forms the base of the throat. 
	  The low energy effective action of a $D3$ brane moving in the warped background (\ref{warped-metric}) is given by Dirac Born Infeld action as follows,
	\begin{equation}
	S=\int d^4 x \sqrt{-g}\left[\frac{R}{2 \kappa^2}+{\cal{P}}(X,\phi)\right]  \label{action-1}
	\end{equation}
	where  \cite{Tong}
	\begin{equation} 
	{\cal{P}}(X,\phi)=- \displaystyle \frac{1}{f(\phi)}\left(\sqrt{1-2f(\phi)X}-1\right)-V(\phi), \qquad X=- \frac{1}{2} g^{\mu \nu}\partial_{\mu}\phi \partial_{\nu}\phi    \label{DBI-part}
	\end{equation}
	 Thus the DBI action on a D3 brane consists of non-canonical kinetic term of the scalar field plus a potential which arises from the moduli stabilization mechanisms and from brane interactions \cite{KKLMMT, Bau}.
	Phenomenologically the action belongs to a general class of models described by non-canonical kinetic term first proposed by 
	Armendariz-Picon et.al as k-inflation models \cite{Mukhanov1} in four dimensions.
In the D-brane scenario, the universe is assumed to be located on the D3 brane where Standard model particles exist in the form open string modes and gravity propagates in all extra directions. Then inflation in the DBI scenario can be realized as a consequence of the relativistic motion of the D3 brane in the higher dimensional spacetime such that all extra six dimensions are compactified and stabilized by flux compactification leading to warped geometry \cite{KKLMMT}. 
The warped brane tension on the D3 brane is given by $f(\phi)^{-1}=\displaystyle \frac{T_{3}}{h^{-1}(\phi)}$ and the inflaton field is related to brane tension $T_3$ as $\phi= \sqrt{T_3} r$. For simplicity, if we assume the D3-brane moves only along the radial direction in the higher dimensional throat neglecting the dilaton effects and other two-form field pullback contributions on the D3 brane, then its motion in the warped spacetime appears as a scalar field from the four dimensional point of view. This single scalar degree of freedom can be interpreted as the inflaton field $\phi(t)$ responsible for accelerated expansion of the early universe.
 The novel feature of the DBI scenario is that due to the presence of non-standard kinetic term and the Lorentz factor $\gamma$, inflation can occur even with steep potentials in contrary to potential required to undergo slow roll inflation. 
In the limit $f(\phi) \rightarrow 0$, the DBI action reduces to the canonical action given by ${\cal{P}}(X,\phi)= X-V(\phi)$.
With the homogeneous, isotropic and spatially flat FRW metric (\ref{metric-1}), the gravitational equations of motion derived using (\ref{action-1}) and (\ref{DBI-part}) are given by,
\begin{eqnarray}
	H^2 =\displaystyle \left(\frac{\dot{a}}{a}\right)^2 & = & \displaystyle\frac{\kappa^2}{3}\left[\frac{\gamma-1}{ f(\phi)}+ \displaystyle V(\phi)\right]  \label{ttcomp} \\[3mm]
	\displaystyle H^2+2\frac{\ddot{a}}{a}&=
	& \displaystyle \kappa^2\left[\frac{1-\gamma}{f(\phi)\gamma}+ V(\phi)\right] \label{iicomp}
\end{eqnarray}
where the Lorentz factor is defined as,
\begin{equation}
{\gamma=\frac{1}{\sqrt{1-f(\phi)\dot{\phi}^2}}}    \label{1}
\end{equation}
since $\dot{\phi}^2$ is associated with the motion of the $D3$ brane, the Lorentz factor $\gamma$ puts constraint on the brane motion in the warped spacetime and sets the limit $f(\phi)\dot{\phi}^2 <1$ in analogy to the special relativistic constraints imposed on a moving particle \cite{Tong}.
From (\ref{ttcomp}) and (\ref{iicomp}), the energy density and pressure can be read off as,
\begin{equation}
\rho  =\frac{\gamma}{f}+(V-f(\phi)^{-1}) ,\qquad
p  =-\frac{1}{f \gamma}-(V-f(\phi)^{-1})   \label{matter1}
\end{equation}
and we can express,
\begin{equation}
\frac{2\dot{H}}{\kappa^2}=\frac{1-\gamma^2}{f\gamma} = -(\rho+p)  \label{2}
\end{equation} 
 If the pressure and the energy density are related as $p= \omega \rho$, then equation of the state parameter $\omega(\phi)$ is given by,
\begin{eqnarray}
\omega(\phi)= \displaystyle \frac{p}{\rho} = \displaystyle \frac{f^{-1}(1-\gamma^{-1})-V(\phi)}{(\gamma-1)f^{-1}+V(\phi)}
\end{eqnarray}
Similar to k-inflation model in general, the speed of sound  $c_s$ represents the propagation speed of perturbations to the inflaton field $\delta \phi$ travelling relative to the homogeneous and isotropic background spacetime which is given by,
\begin{eqnarray}
c_s &=& \sqrt{\displaystyle \frac{dp}{d\rho}}
\end{eqnarray}
and in general is a time dependent quantity related to the Lorentz factor $\gamma$ as, 
\begin{equation}
\gamma =\displaystyle \frac{1}{c_s} 
\end{equation}
The equation of motion of the inflaton field in the FRW background is given by, 	
\begin{equation}
\ddot{\phi}+\left(\frac{3f'}{2f}\right)\dot{\phi}^2-\frac{f'}{f^2}+\frac{3 H \dot{\phi}}{\gamma^2}+\frac{1}{\gamma^3} \left(V'+\frac{f'}{f^2}\right)=0 \label{scalar}
\end{equation}
which can also be expressed in the following form,
\begin{equation}
\frac{d}{dt}(\gamma \dot{\phi})+3H\dot{\phi}\gamma+ V'=0
\end{equation} 
It can be readily seen that the canonical case is obtained in the limit $\gamma \rightarrow 1$ implying $c_s=1$.
In analogy to the slow roll parameters defined for canonical inflation, the flow parameters in DBI theory can be defined as follows,
\begin{eqnarray}
\epsilon_{H} &=& -\frac{\dot{H}}{H^2}  \label{ep-H} \\
	\eta_{H} &=& \frac{\dot{\epsilon_{H}}}{\epsilon_{H} H} \label{eta-H} \\
s_{H} &=& \frac{\dot{c_s}}{c_s H}	
\end{eqnarray} 
The additional parameter $s_{H}$ arises due to varying sound speed, a feature which is shared by all k-inflation models. Therefore the curvature perturbations of DBI inflation models in particular are influenced by the sound horizon $ kc_s=aH$.

\section{Constant-roll inflation with invariable speed of sound} 
In the present paper we generalize the constant roll inflation model by taking non-standard kinetic term specifically in the context of DBI scalar field theories. 
We aim to determine exact inflationary solutions by employing Hamilton-Jacobi formalism in the DBI scenario when the $D3$ brane undergoes relativistic motion in a given throat geometry.  
Considering the scalar field $\phi$ acts as a time variable such that the Hubble parameter becomes a function of $\phi$,
we have $\dot{H}= H'(\phi)\dot{\phi}$. Then using (\ref{1}) and (\ref{2}) we get,
\begin{equation}
 \dot{\phi}=-\frac{2}{\kappa^2}\frac{H'(\phi)}{\gamma}  \label{3}
\end{equation}
We note that the above relation can also be obtained by differentiating (\ref{ttcomp}) with respect to time and subsequently substituting in it (\ref{scalar}). Using  (\ref{1}) and (\ref{3}), one can express $\gamma$ and $\dot{\phi}$ purely as a function of $\phi$ as follows,
\begin{equation}
\gamma(\phi)=\sqrt{1+\frac{4 f(\phi) H'(\phi)^2}{\kappa^4}}  \label{gamma}
\end{equation}
so that,
\begin{equation}
\dot{\phi}=\frac{-2H'}{\sqrt{\kappa^4+4 f H'^2}}  \label{phi_dot1}
\end{equation}
and the warp factor is given by,
\begin{equation}
f(\phi)= \displaystyle \frac{\kappa^4[\gamma^2(\phi)-1]}{4 H'(\phi)^2}  \label{gamma-1}
\end{equation}
Substituting (\ref{gamma}) in (\ref{ttcomp}), the Hamilton-Jacobi equation becomes,
\begin{equation}
\frac{3 H(\phi)^2}{\kappa^2}-\frac{1}{f(\phi)}
\sqrt{1+\frac{4f(\phi )H'(\phi)^2}{\kappa^4}}+\frac{1}{f(\phi)}=V (\phi) \label{HJeqn1}
\end{equation} 
 In general flow parameters in DBI inflation in terms of the derivatives of $H(\phi)$ can be defined as follows,
 \begin{eqnarray}
 \epsilon(\phi)&=&\frac{2}{\kappa^2}\frac{1}{\gamma(\phi)}\left(\frac{H'}{H}\right)^2  \label{ep-phi} \\
 \eta(\phi)&=&\frac{2}{\kappa^2}\frac{1}{\gamma(\phi)}\left(\frac{H''}{H}\right)   \label{eta-phi} \\
 \xi(\phi)&=& \frac{4}{\kappa^2} \frac{1}{\gamma^2(\phi)}\left(\frac{H' H''}{H^2}\right) \label{xi-phi} 
 \end{eqnarray}
Similarly, higher order flow parameters can be constructed \cite{Peris}. 
 In this case, the e-folding number can be defined as,
 \begin{equation}
 N= -\int H dt = \frac{\kappa}{\sqrt{2}} \int^{\phi_i}_{\phi_{f}} \sqrt \frac{\gamma(\phi)}{\epsilon(\phi)} 
 \end{equation}
 where $\phi_{f}$ is the value of inflaton field at the end of inflation.
Due to the existence of a variable sound speed additional flow parameters emerge and quantify variation of the sound speed $c_s$ with respect to $\phi$ which are defined as follows,
 \begin{equation}
 s_{1}(\phi)=\frac{2}{\kappa^2 \gamma(\phi)} \frac{\gamma' H'}{\gamma H},  \qquad
 s_{2}(\phi)=\frac{2}{\kappa^2 \gamma(\phi)} \frac{\gamma''}{\gamma} 
 \end{equation}
 An hierarchy of sound flow parameters can be constructed in analogy to Hubble flow parameters. The flow parameters defined with respect to time and inflaton field are related
  where it can shown that $\epsilon_{H}=\epsilon$ and $\eta_{H}=2\epsilon-2\eta-s$. 
  The equation of motion of the inflaton field can be expressed in terms of the flow parameters as follows \cite{Kinney1},
 \begin{equation}
 2 \displaystyle \left(\frac{\gamma}{\gamma +1}\right)\eta - \left(\displaystyle \frac{\gamma}{\gamma+1}\right)^2 s_1-3-\displaystyle \frac{V'}{H \dot{\phi} \gamma}=0   
 \end{equation}
 In the same spirit of canonical constant roll inflation, the corresponding non-canonical generalization of the constant roll inflation for the particular case of the scalar DBI theory can be expressed as,
 \begin{equation}
 2 \displaystyle \left(\frac{\gamma}{\gamma +1}\right)\eta - \left(\displaystyle \frac{\gamma}{\gamma+1}\right)^2 s_1= (3+ \alpha_D)  \label{6}
 \end{equation}
 where we define, 
\begin{equation}
 \alpha_{D}=\displaystyle \frac{V'}{H \dot{\phi} \gamma}= \displaystyle \frac{\alpha}{\gamma}
 \end{equation}
 $\Rightarrow$
 \begin{equation}
 \alpha_D=c_s \alpha
 \end{equation}
 where $\alpha$ is the constant roll parameter in the canonical case defined by (\ref{alpha-can}).
 The parameter $\alpha_D$ is the DBI analog of the constant roll inflation parameter defined in the canonical set-up. Being proportional to $c_s$, $\alpha_D$ is not a spacetime independent quantity like $\alpha$ and hence presents deviation from inflationary scenarios described by exactly flat potential with the DBI action which specifically implies $\alpha_{D}=0$.
 It may be noted that $\alpha_D$ reduces to its canonical counterpart $\alpha$ and becomes constant parameter when  $f(\phi) \rightarrow 0$ i,e. $\gamma= c_s=1$.
Assuming the equation of state as $p = \omega(\phi) \rho$, $\omega(\phi)$ can be determined using (\ref{ttcomp}), (\ref{matter1}) and (\ref{gamma}) as follows, 
\begin{equation}
\displaystyle \frac{4 H'^2}{\kappa^2}=3(\omega +1)H^2 \gamma
\label{EOS}
\end{equation}
\subsection{Cosmological solutions}
In the present paper, we shall investigate family of inflationary solutions with constant $\gamma$ i,e. $c_s$  is constant $c_s \neq 1$, thus keeping ourselves in the DBI regime by considering and at the same time restricting our study to the simplest modification to the canonical case that corresponds to $\gamma=c_s=1$.
With this consideration, all sound flow parameters vanish i,e.  $s_{1}=s_{2}=0$ and $\alpha_D$ becomes a constant quantity as a result the second order differential equation (\ref{6}) is significantly simplified as, 
\begin{equation}
H''=\frac{\kappa^2}{2}\left(\frac{\gamma+1}{2\gamma}\right)(3+\alpha_{D})H \gamma   \label{7}
\end{equation}
which exactly matches with (\ref{Hubb-can}) in the canonical limit and can be analytically solved to obtain solutions as,
\begin{equation}
H(\phi)=C_1 e^{\frac{1}{2}\sqrt{(\gamma +1)(3+\beta)} \kappa  \phi}+C_2 e^{-\frac{1}{2}\sqrt{(\gamma +1)(3+\beta)} \kappa  \phi}  \label{8}
\end{equation}
where $C_1$, $C_2$ are constants of integration. 
Without the loss of generality, in the constant $\gamma$ set-up, we replaced $\alpha_{D}$ by the parameter $\beta$ and $\gamma$ by a constant positive parameter as,
\begin{equation}
\gamma  =m \label{gamma-2}, \qquad c_{s}=\displaystyle \frac{1}{m}
\end{equation}
such that $m \geq 1$ is always satisfied. Then using (\ref{gamma-2}), the warp factor can be expressed as,
\begin{equation}
f(\phi)= \displaystyle \frac{\kappa^4}{4 H'^2} \left(m^2-1\right) \label{warp-general}
\end{equation}
 Using the Hamilton-Jacobi equation (\ref{HJeqn1}) and (\ref{gamma-2}), the inflaton potential is found to be,
\begin{equation}
V(\phi)=\frac{3 H^2}{\kappa^2}-\frac{4H'^2}{\kappa^2} \left(\frac{1}{m+1}\right)  \label{HJeqn2}
\end{equation}  
and the equation of state parameter can be expressed using (\ref{EOS}) as,
\begin{equation}
\omega(\phi)=-1+\frac{4}{3 \kappa^2 m}\left(\frac{H'}{H}\right)^2  \label{EOS1}
\end{equation}
In general, $\omega(\phi)$ is a slowing varying function of $\phi$, which is only if $\displaystyle \frac{H'(\phi)}{H(\phi)}$ remains constant throughout the inflationary phase.
Now (\ref{8}) gives rise to two different classes of solutions of $H(\phi)$ corresponding to two parameter regions of the DBI constant roll parameter,
 namely $\beta >-3$ and $\beta <-3$.
We shall investigate each case separately in this section. 
\begin{itemize}
	\item $\gamma=\displaystyle m$, $\beta>-3$ \\[1mm]
In this parameter region, there exist three independent solutions of $H(\phi)$.  
After a little rearrangement of the constants of integration,  one finds that (\ref{8}) admits an independent solution consisting of purely exponential function given by,
\begin{eqnarray}
H(\phi)=C e^{ \frac{1}{2} \sqrt{(3+\beta)(m+1)} \kappa\phi } \label{Hubble_exp}
\end{eqnarray}
where $C$ is a constant. 
As the exponential solution of $H(\phi)$ is symmetric under $\phi \rightarrow -\phi$, we take the positive root only. Substituting (\ref{Hubble_exp}) in (\ref{warp-general}), the warp factor is given by,
\begin{equation}
f(\phi)= \frac{\kappa^2(m-1)}{ C^2  (3 + \beta)}e^{- \sqrt{(3+\beta)(m+1)}  \kappa \phi }
\end{equation}
which is an exponentially decaying function of $\phi$ as seen from the plot of $f(\phi)$ vs $\phi$ in the Fig. 1 and becomes vanishingly small in the large $\phi$ limit thus indicating that the influence of the non-standard kinetic term diminishes as $\phi$ becomes large.
\begin{figure}
	\begin{minipage}{0.4\textwidth }
	\includegraphics[width=2.5 in]{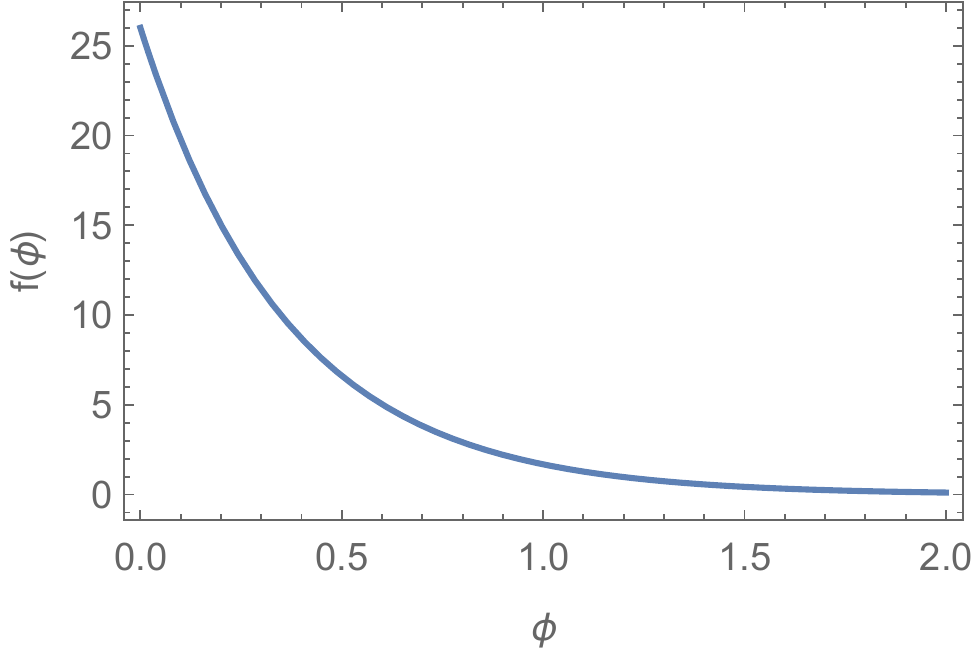}
	\caption{$f(\phi)$ vs $\phi$ plot}
	\end{minipage}
	\hfill
	\begin{minipage}{0.5\textwidth }
		\includegraphics[width=2.5 in]{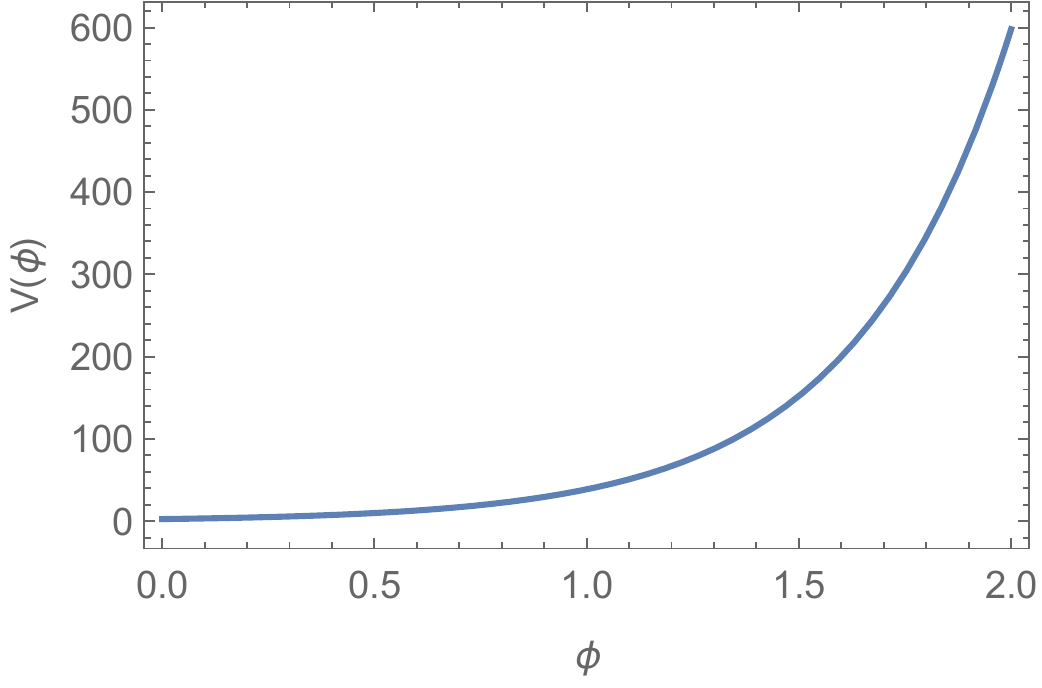}
		\caption{Plot of $V(\phi)$ vs $\phi$.}
	\end{minipage}
\caption*{ With $\beta=-2.5$, $m=14$, $C=1$ and $\kappa=1$ for power-law inflation in the region $\beta>-3$.}
\end{figure}
Using (\ref{HJeqn2}) the inflaton potential is obtained as follows,
\begin{equation}
V(\phi)=-\frac{C^2}{\kappa^2} \beta e^{ \sqrt{(3+\beta)(m+1)} \kappa \phi} \label{pot-expo}
\end{equation}  
which is positive for any $\beta<0$ and is real provided $3+\beta>0$ as shown in Fig 2. Such a form of inflaton potential mimics the power law inflation.  In the DBI set-up, Spalinski \cite{Spalinski1} first studied power law inflation in the ultra relativistic limit with $\dot{c_s}=0$.
\begin{figure}
	\begin{minipage}{0.5\textwidth }
		\includegraphics[width=2.3 in]{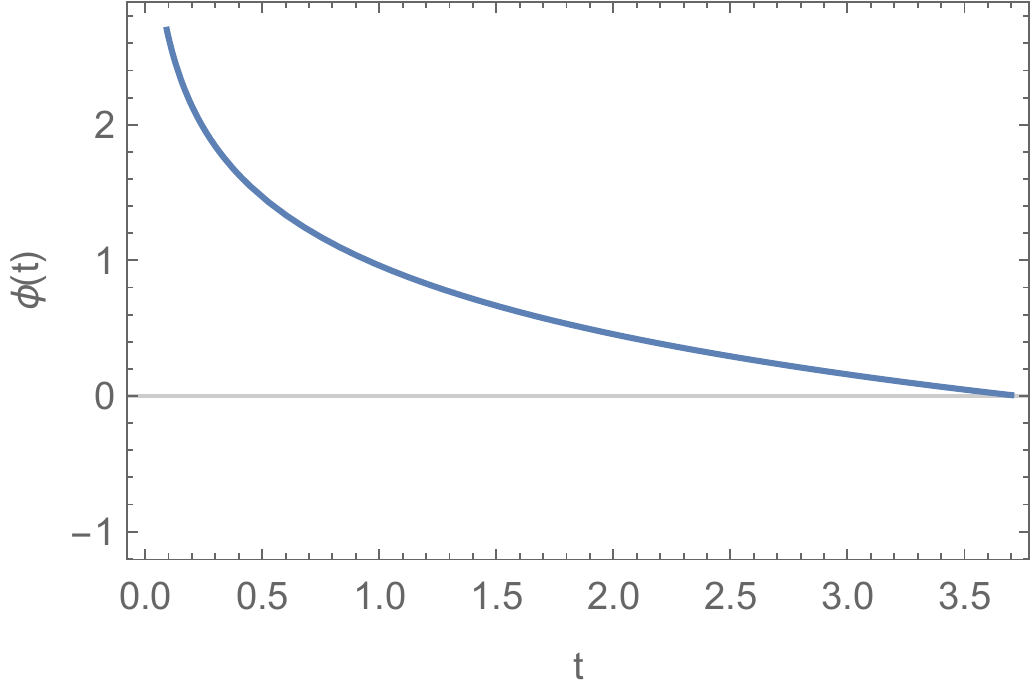}
		\caption{ $\phi(t)$ vs $t$ plot.}
			\end{minipage}
		\hfill
		\begin{minipage}{0.5 \textwidth}
	\includegraphics[width=2.3 in]{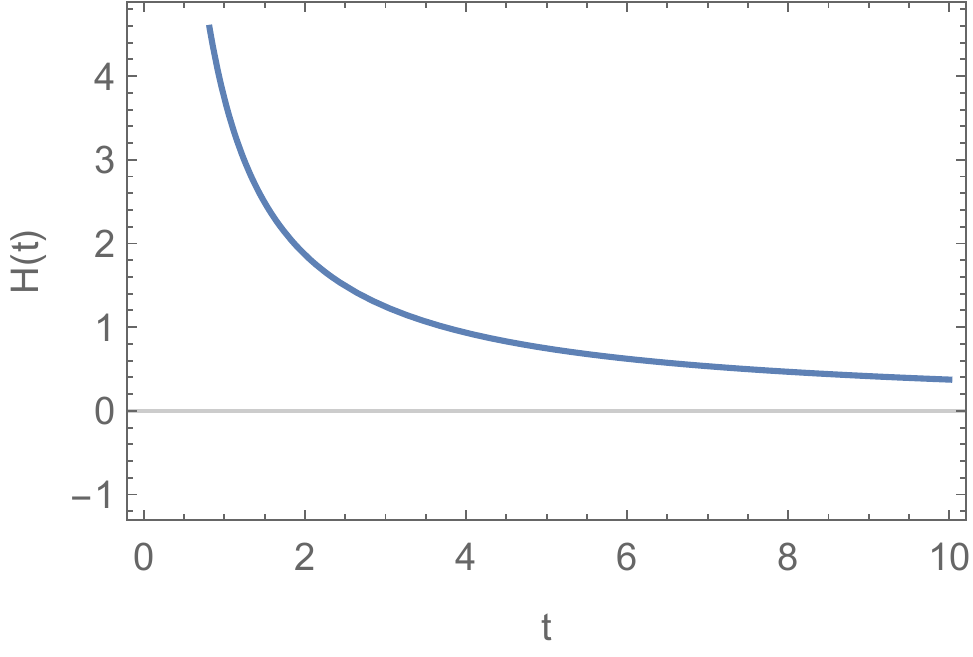}
	\caption{Plot of $H(t)$ vs $t$ .}
	\end{minipage}
\caption*{For the choice of parameters $\beta=-2.5$, $m=14$, $C=1$ and $\kappa=1$. }
\end{figure}
The scalar field profile is determined by using (\ref{3}) and (\ref{Hubble_exp}) as,
\begin{equation}
\phi(t) = - \frac{2}{\kappa\sqrt{ (3+\beta)(m+1)}} \ln \left[\frac{(3+\beta)(m+1)}{2m}C t \right]
\end{equation}
The time evolution of the scalar field profile during the power law inflation is depicted in Fig. 3. Substituting $\phi(t)$ back in $H(\phi)$ the Hubble parameter for the power law inflation is obtained as,
\begin{equation}
H(t)=\displaystyle \frac{2m}{(3+\beta)(m+1)t}
\end{equation}
which decreases with time and eventually becomes constant as shown in Fig. 4. The corresponding scale factor the DBI power law inflation is given by,
\begin{equation}
a(t) = a_0 t^{\frac{2m}{(3+\beta)(m+1)}}
\end{equation}
where $a_0$ is the constant of integration.  
From (\ref{EOS1}), the equation of state parameter  $\omega(\phi)$ is constant in $\phi$ as the ratio $\displaystyle \frac{H'(\phi)}{H'(\phi)}$ remains constant during the power law inflation. So for the power law case we have, 
\begin{equation}
\omega=\displaystyle \frac{3+\beta(m+1)}{3m}
\end{equation}
The power law inflation in the DBI scenario is strongly constrained by observations due to its prediction of large non-Gaussianity and enhanced tensor perturbations. 
However, recently it has been showed that DBI power law inflation in the framework of constant speed of sound might survive in the slow roll regime provided $c_s\geq 0.087$  is always satisfied \cite{Amani}.
The remaining two independent solutions in the parameter region $\beta>-3$ are constructed from hyperbolic functions. 
The cosine hyperbolic solution of $H(\phi)$ is given by,
\begin{equation}
H(\phi)=C \cosh\left(\frac{1}{2}\sqrt{(m+1)(3+\beta)} \kappa \phi \right)
 \label{Hubble2}
\end{equation}
and the corresponding solution of the warp factor becomes,
\begin{equation}
	f(\phi)=\frac{\kappa ^2 (m-1) \text{csch}^2\left(\frac{1}{2}  \sqrt{(\beta +3) (m+1)}\kappa  \phi \right)}{C^2 (\beta +3)}
\end{equation}
which exactly vanishes with $m=1$ in the canonical limit.
In contrast to the power law inflation, the hyperbolic solution permits $\beta>0$ as well as $\beta<0$ provided $3+\beta>0$ is satisfied.
The variation of the warp factor with the inflaton field depicted in fig. 5 for both $\beta>0$ and $\beta<0$ shows that the behavior of the warp factor is insensitive to the sign of $\beta$ and exponentially converges to the canonical limit for large values of $\phi$.
\begin{figure}[h]
\begin{minipage}{0.5\textwidth}
	\includegraphics[width=2.3 in]{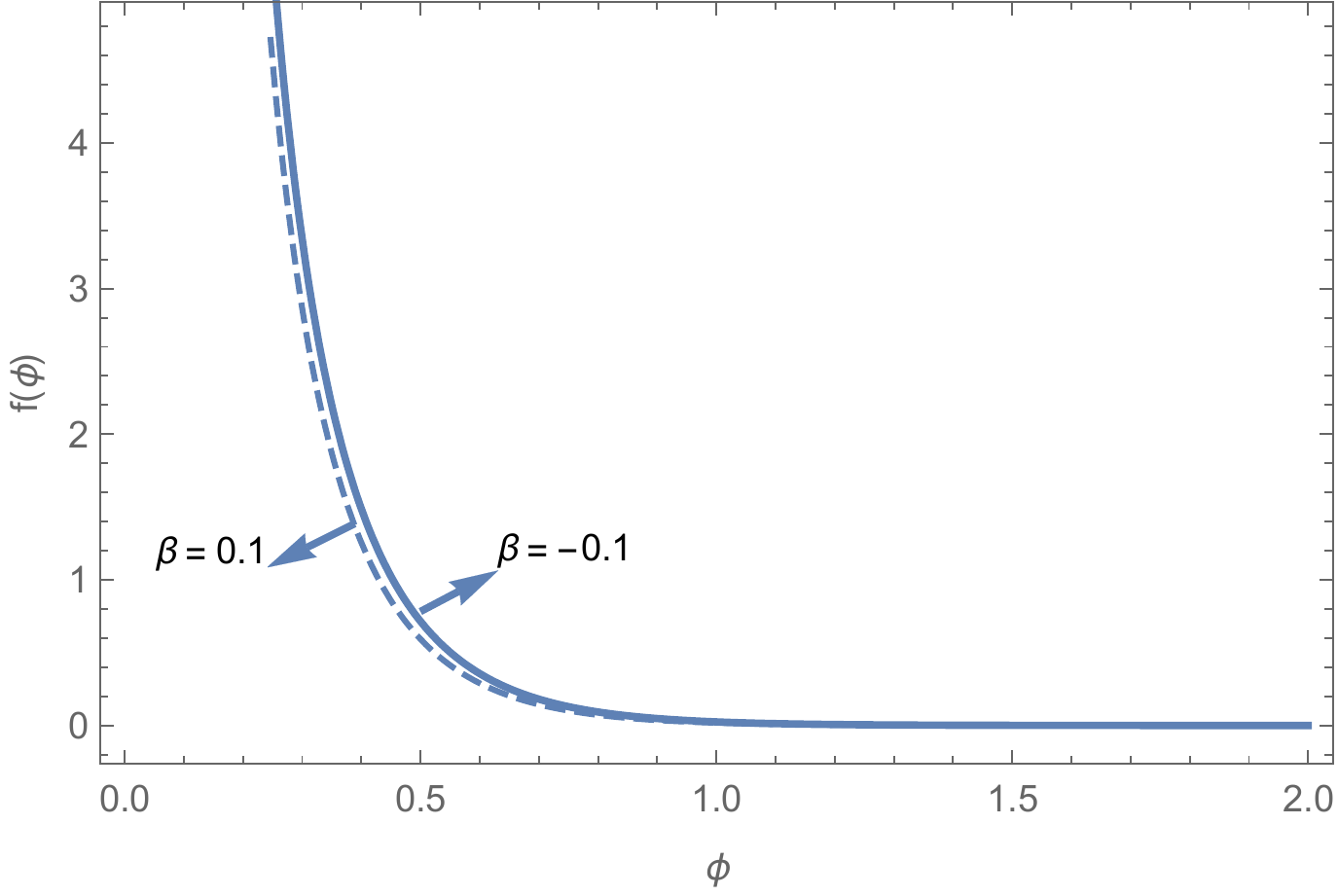}
	\caption{$f(\phi)$ vs $\phi$ plot}
\end{minipage}
\hfill
\begin{minipage}{0.5 \textwidth}
\includegraphics[width=2.3 in]{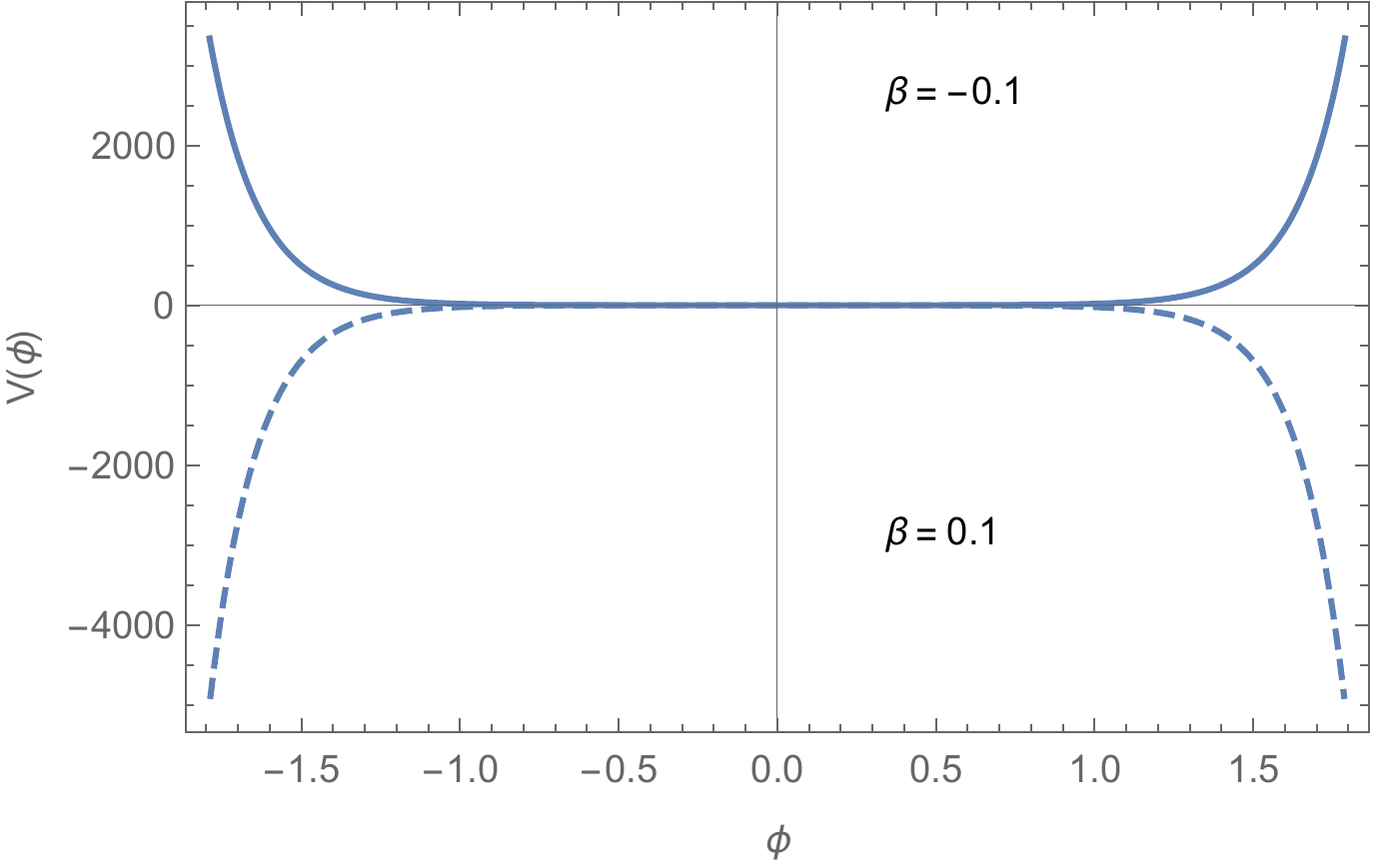}
\caption{ $V(\phi)$ vs $\phi$ plot.}
	\end{minipage}
\caption*{Values of parameters : $m=14$, $C=1$ and $\kappa=1$. }
\end{figure}
In this case, the inflaton potential is given by, 
\begin{eqnarray}
V(\phi)&=& \displaystyle \frac{3C^2}{ \kappa^2} \left[1+\frac{\beta}{6}\left\lbrace 1-\cosh \left(\sqrt{(m+1)(3+\beta)}\kappa \phi\right)\right\rbrace \right]  
\label{cosh-V}
\end{eqnarray}
which is minimum given by $V(\phi)= \displaystyle \frac{3C^2}{\kappa^2}$ at $\phi=0$ and exactly reduces to canonical form for $m=1$ \cite{Starobinsky1}.
In particular $\beta=0$ corresponds to inflationary scenario with flat potential given by $V(\phi)= \displaystyle \frac{3C^2}{\kappa^2}$. The inflationary solutions with constant potential in the canonical context known as the ultra-slow roll inflation has been studied \cite{ultra-roll, Kinney1} and the more general case  involving the DBI theory was studied by Spalinski \cite{Spalinski1}. 
The fig. 6 depicts the variation of the inflaton potential with respect to the inflaton field which shows that irrespective of the value of $m$, the inflation potential always remains positive for $\beta<0$ while $\beta>0$ leads to an inverted negative potential.
In this case, the scalar field profile, Hubble parameter and the scale factor are then given by,
\begin{eqnarray}
\phi(t)&=& \displaystyle\frac{1}{\kappa} \frac{2}{\sqrt{(m+1)(3+\beta)}} \ln \left[\coth \left(\frac{(m+1)(3+\beta)}{2m}C t\right)\right]\\[2mm]
H(t)&= & C \coth[\frac{(m+1)(3+\beta)}{2m} C t] \\[2mm]
a(t)&= &a_1 \sinh^{2/(3 +\beta)(m+2)}[\frac{(m+1)(3+\beta)}{2m} C t ]
\end{eqnarray}
where $a_1$ is a proportionality constant.
The time evolution of the scale field profile is shown in Fig. 7 and Fig. 8 shows that Hubble parameter remains almost constant during inflation characterized by the inflaton potential (\ref{cosh-V}).
 \begin{figure}[h]
	\begin{minipage}{0.5\textwidth }
		\includegraphics[width=2.5 in]{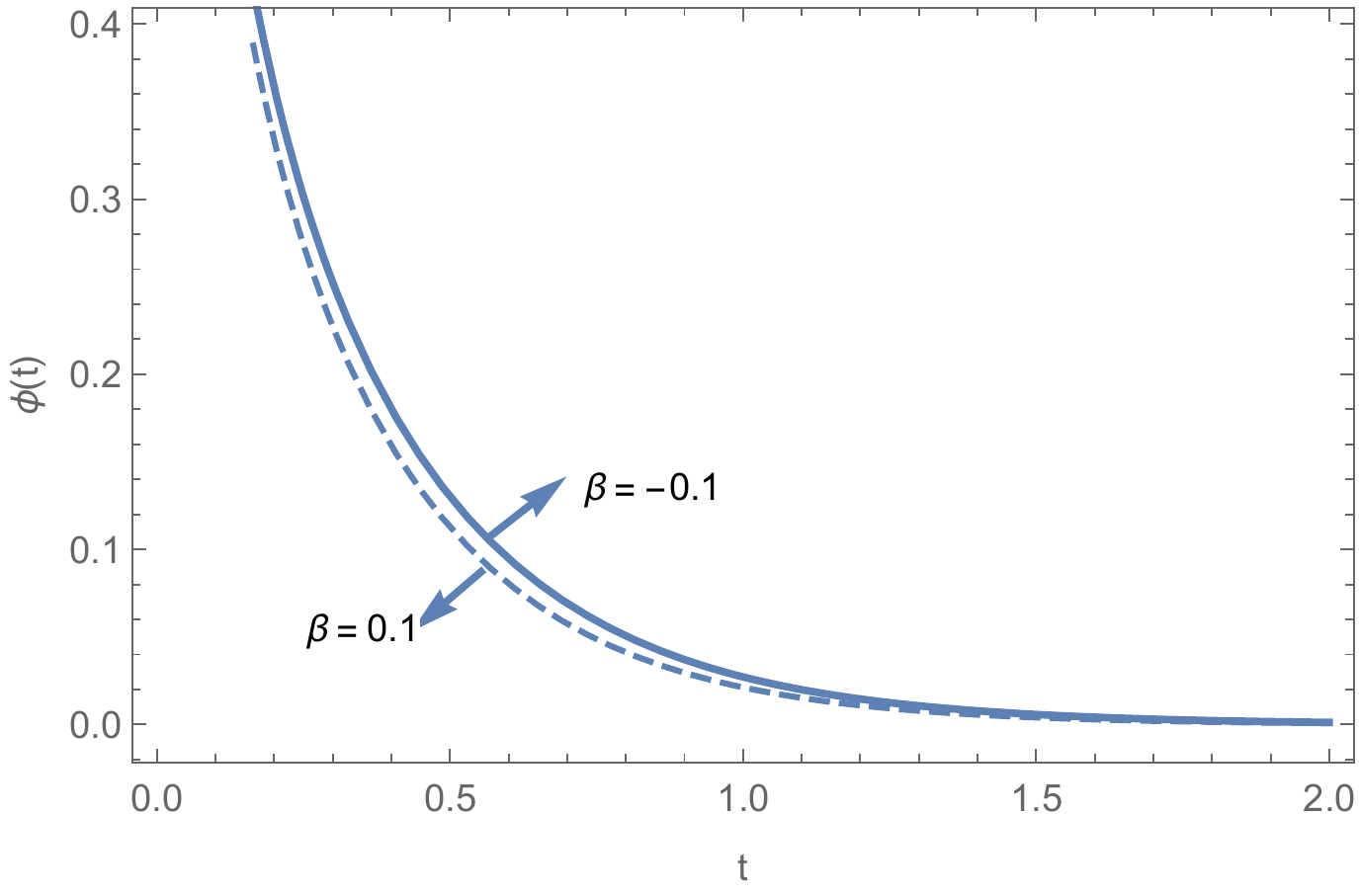}
		\caption{Plot of $\phi(t)$ vs $t$.}
	\end{minipage}
	\hfill
	\begin{minipage}{0.5 \textwidth}
		\includegraphics[width=2.5 in]{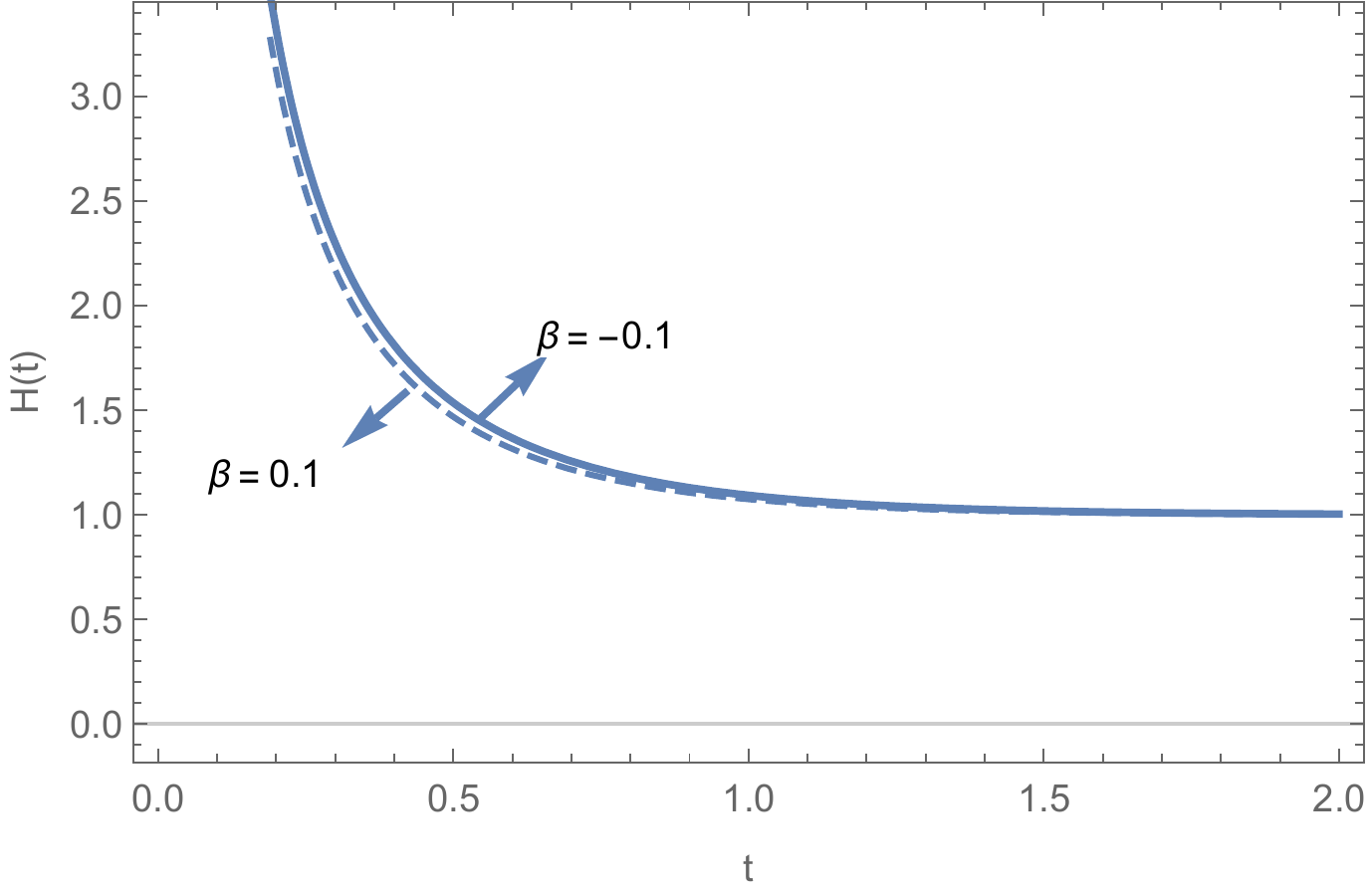}
		\caption{$H(t)$ vs $t$ plot.}
	\end{minipage}
	\caption*{Values of Parameters: $m=14$, $C=1$ and $\kappa=1$. }
\end{figure}
 The equation of state parameter for the given Hubble function is given by,
\begin{equation}
w(\phi)= -1+\frac{(\beta +3) (m+1) \tanh ^2\left(\displaystyle \frac{1}{2} \sqrt{(\beta +3) (m+1)}\kappa  \phi\right)}{3 m}
\end{equation}
which approaches $-1$ in the small $\phi$ limit and becomes constant in the large $\phi$ region. As $\phi$ increases, $\omega(\phi)$ increases from -1 and finally attains a constant positive value for $\beta\geq0$. 
For $\beta<0$, $\omega(\phi)<0$ in the large $\phi$ limit  but becomes positive as $\beta \rightarrow 0$. This behaviour of $\omega(\phi)$ closely resembles the evolution of the universe from inflation to the radiation-dominated phase ($\omega=0$) and the matter dominated phase ($\omega>0$). 
 For the given inflaton potential (\ref{cosh-V}), the Hubble flow parameters  $\epsilon_{H}$  and $\eta_{H}$ are obtained as follows,
\begin{equation}
	\epsilon_{H}=\frac{(3+\beta) (m+1)}{m(\cosh [ \frac {(3+\beta) (m+1) Ct}{m}]+1)}, \quad \eta_{H}=-\frac{(3+\beta ) (m+1) \tanh ^2\left(\frac{ (3+\beta ) (m+1)C t}{2 m}\right)}{m}
\end{equation}
such that at late times i,e. $Ct >>1$, $\epsilon_{H} \rightarrow 0$ as shown in Fig. 9. The time evolution of $\eta_{H}$ given in Fig. 10 shows that $\eta_{H}$ is non-negligible at late times for a given $m$ which precisely implies the breakdown of slow roll approximations in the DBI regime subjected to constant roll conditions.  A similar conclusion can be inferred using DBI flow parameters $\epsilon(\phi)$ and $\eta(\phi)$ expressed using (\ref{ep-phi}) and (\ref{eta-phi}) as follows,
\begin{equation}
\epsilon(\phi)=\frac{(\beta +3) (m+1) \tanh ^2\left(\frac{1}{2} \sqrt{(\beta +3) (m+1)}\kappa  \phi\right)}{2 m}
, \quad \eta(\phi)=\frac{(\beta +3) (m+1)}{2 m}
\end{equation}
In the small field limit, $\epsilon(\phi)$ approaches zero  whereas $\eta(\phi)$ settles to a constant value indicating that although inflation occurs but slow approximations ($\eta_{H}<<1, \eta <<1$) are violated.
\begin{figure}[h]
	\begin{minipage}{0.5\textwidth }
		\includegraphics[width=2.5 in]{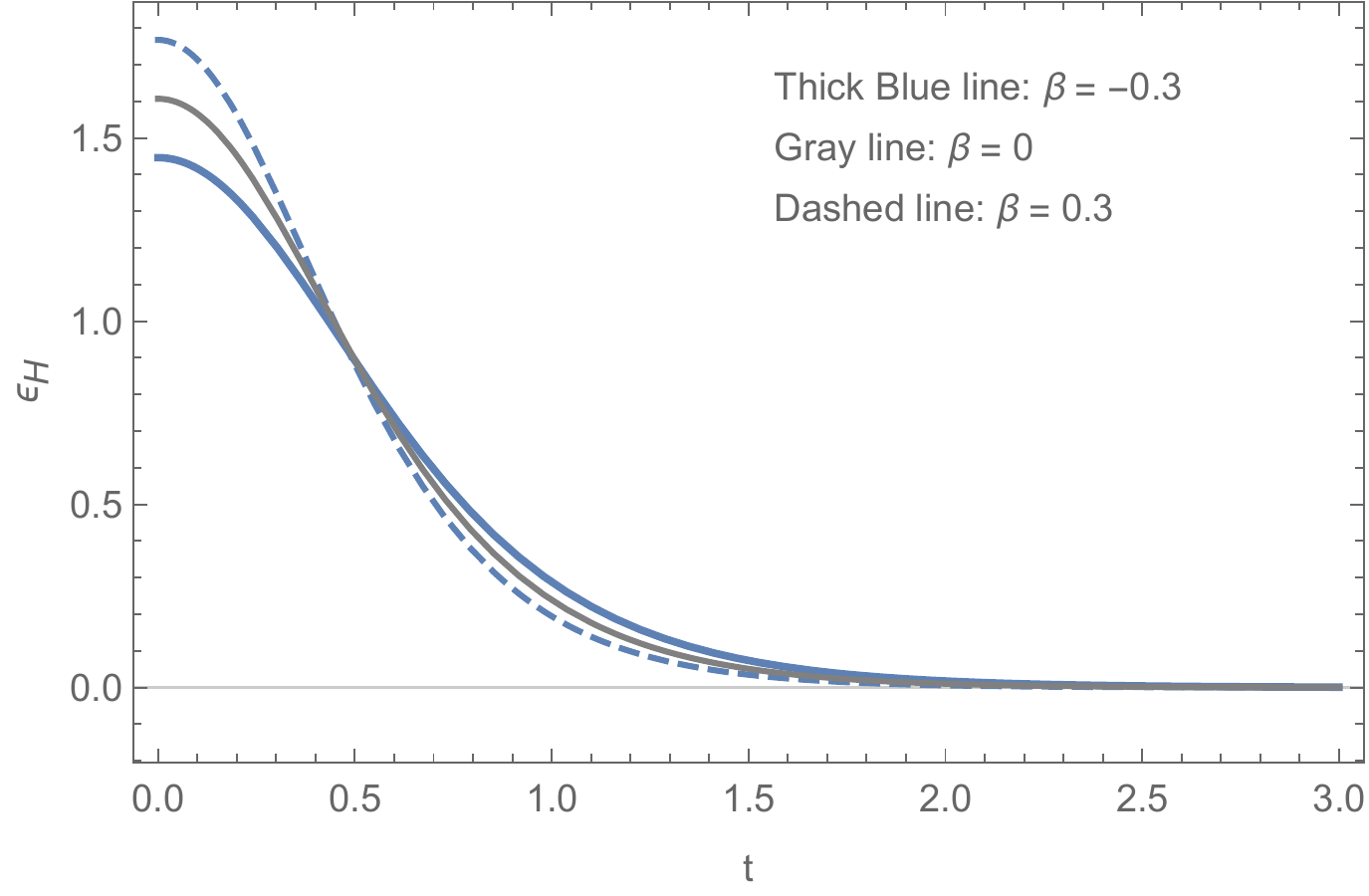}
		\caption{Plot of $\epsilon_{H}$ vs $t$.}
	\end{minipage}
	\hfill
	\begin{minipage}{0.5 \textwidth}
		\includegraphics[width=2.5 in]{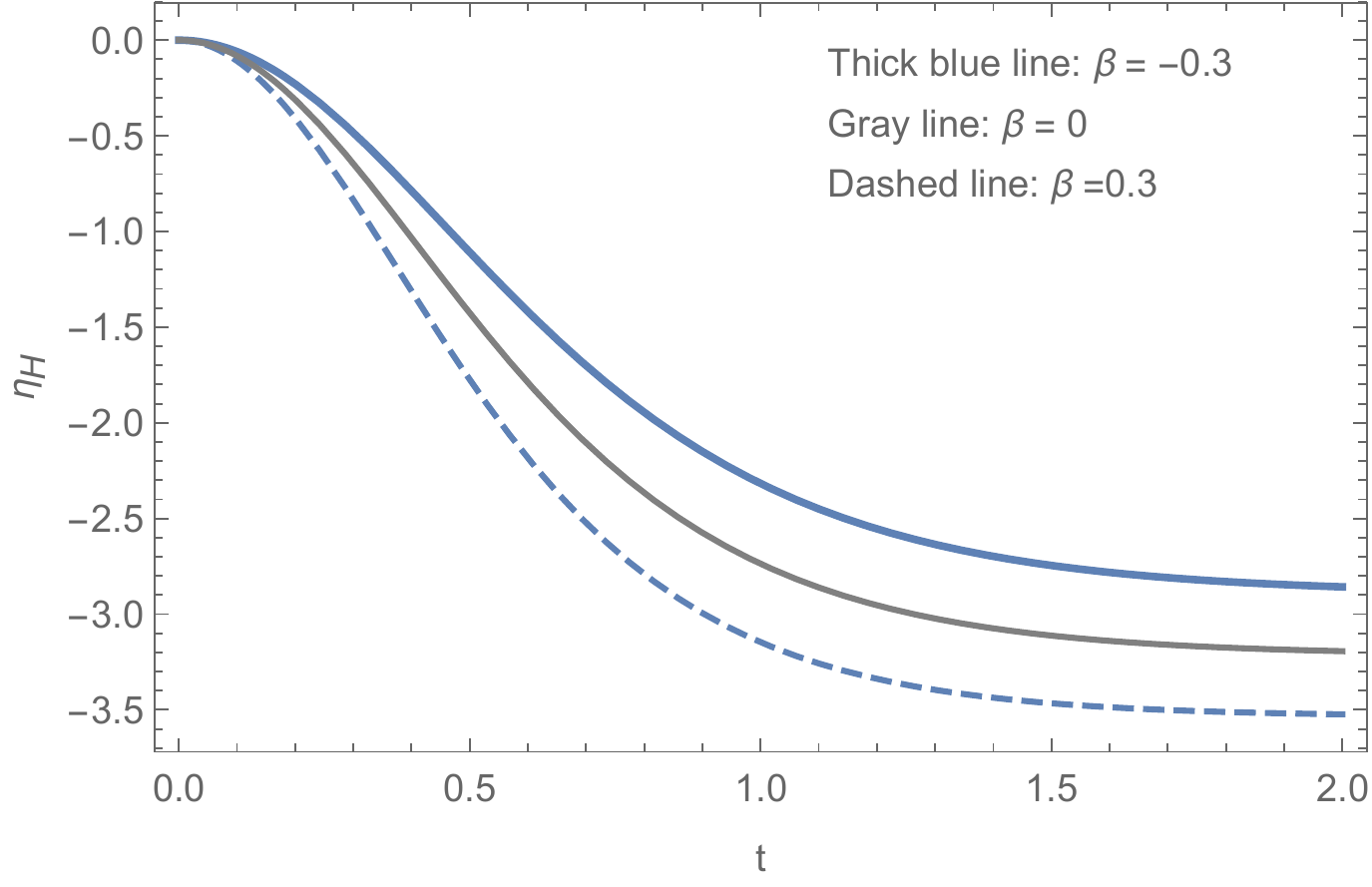}
		\caption{$\eta_H$ vs $t$ plot.}
	\end{minipage}
	\caption*{Choice of parameters: $m=14$, $C=1$ and $\kappa=1$. }
\end{figure}
The third independent solution is given by,
\begin{equation}
H(\phi)=C \sinh\left[\frac{1}{2}\sqrt{(m+1)(3+\beta)} \kappa \phi \right]
\end{equation}
with the corresponding DBI warp factor as,
\begin{equation}
f(\phi)=\frac{\kappa ^2 (m-1) sech^2\left(\frac{1}{2}   \sqrt{(\beta +3) (m+1)}\kappa  \phi\right)}{C^2 (\beta +3)}
\end{equation}
In this case, the inflaton potential, scalar field profile,  Hubble parameter and the scale factor are given by, 
\begin{eqnarray}
V(\phi)&=&-\displaystyle \frac{3C^2}{ \kappa^2}\left[1+\beta\left\lbrace 1+\cosh(\sqrt{(m+1)(3+\beta)})\right\rbrace\right]\\ [2mm]
\phi(t)&=&-\displaystyle \frac{2}{\kappa}\sqrt{\frac{4}{(m+1)(3+\beta)}} \tanh^{-1}{\left[\tan \left(\frac{(m+1)(3+\beta)}{4m}C t\right)\right]}\\[2mm]
	H(t)&=& -C \tan \left[\frac{(m+1)(3+\beta)}{2m}C t \right]\\[2mm]
	a(t)&=&a_2 \cos^{2m/(m+1)(3+\beta)}\left[ \frac{(m+1)(3+\beta)}{2m}C t \right]  \label{scale}
\end{eqnarray}
Unlike the exponential and the cosine hyperbolic solutions, from (\ref{scale}), one finds that $\ddot{a}(t)<0$. Therefore the sine hyperbolic solution does not describe inflationary phase instead the solution indicates a bouncing scenario of the universe. Such a solution also arises in the canonical set-up \cite{Starobinsky1}. 

\item $\gamma=m$, $\beta<-3$ \\[1mm]
In the parameter region $\beta+3<0$, inflationary solutions given by (\ref{8})  with constant $\gamma$ turn out to be trigonometric solutions of $H(\phi)$. But the sine solution is not mathematically permitted because it gives rise to unacceptable imaginary $H(\phi)$. As a result, the only non-trivial and independent solution is given by,
\begin{equation}
H(\phi)=C \cos\left(\frac{1}{2}\sqrt{(m+1)|3+\beta|} \kappa \phi \right)
\label{Hubble3}
\end{equation}
The corresponding warp factor in this case is found to be,
\begin{equation}
f(\phi)= \frac{\kappa ^2 (m-1) \csc ^2\left(\displaystyle\frac{1}{2}   \sqrt{|3+\beta| (m+1)}\kappa  \phi\right)}{ C^2 |3+\beta|}  \label{cos-warp}
\end{equation}
which shows $f(\phi)$ is periodic with periodicity at $\phi_d=\displaystyle \frac{2}{\kappa}\frac{n \pi}{\sqrt{|3+\beta|(m+1)}}$
where $n=0, \pm 1, \pm 2,..$. Moreover, in the large $\phi$ limit, $f(\phi) \sim \exp[-\sqrt{(m+1)(3+\beta)}\kappa \phi]$ as shown in the plot of $f(\phi)$ vs $\phi$ in figure 11.
\begin{figure}
	\begin{minipage}{0.5\textwidth}
\includegraphics[width=2.5 in]{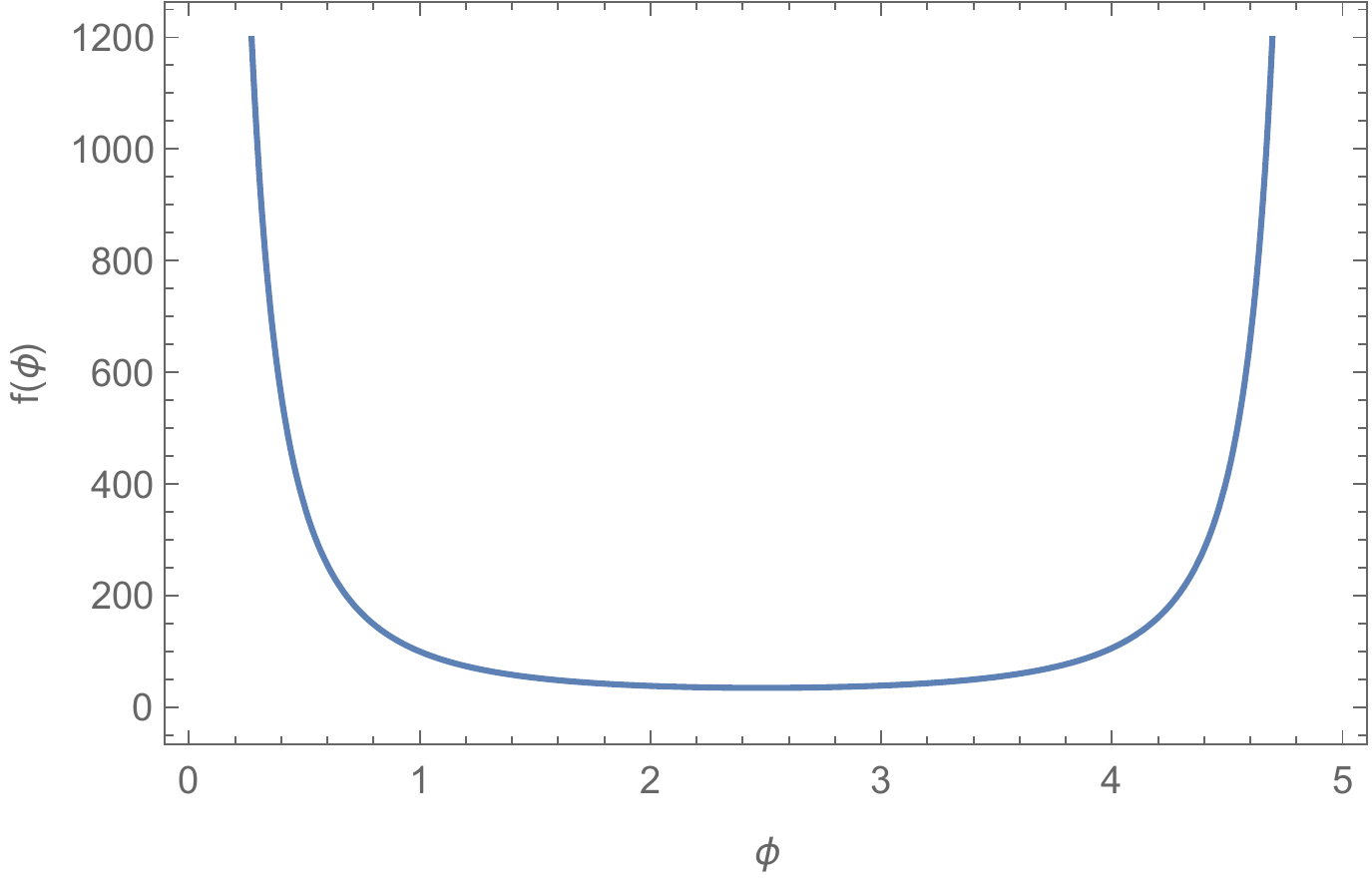}
\caption{$f(\phi)$ vs $\phi$ plot with $m=14,\\ \beta=-3.5,\kappa=1$ and $C=1$.}		
	\end{minipage}
\hfill
\begin{minipage}{0.5 \textwidth}
	\includegraphics[width=2.5 in]{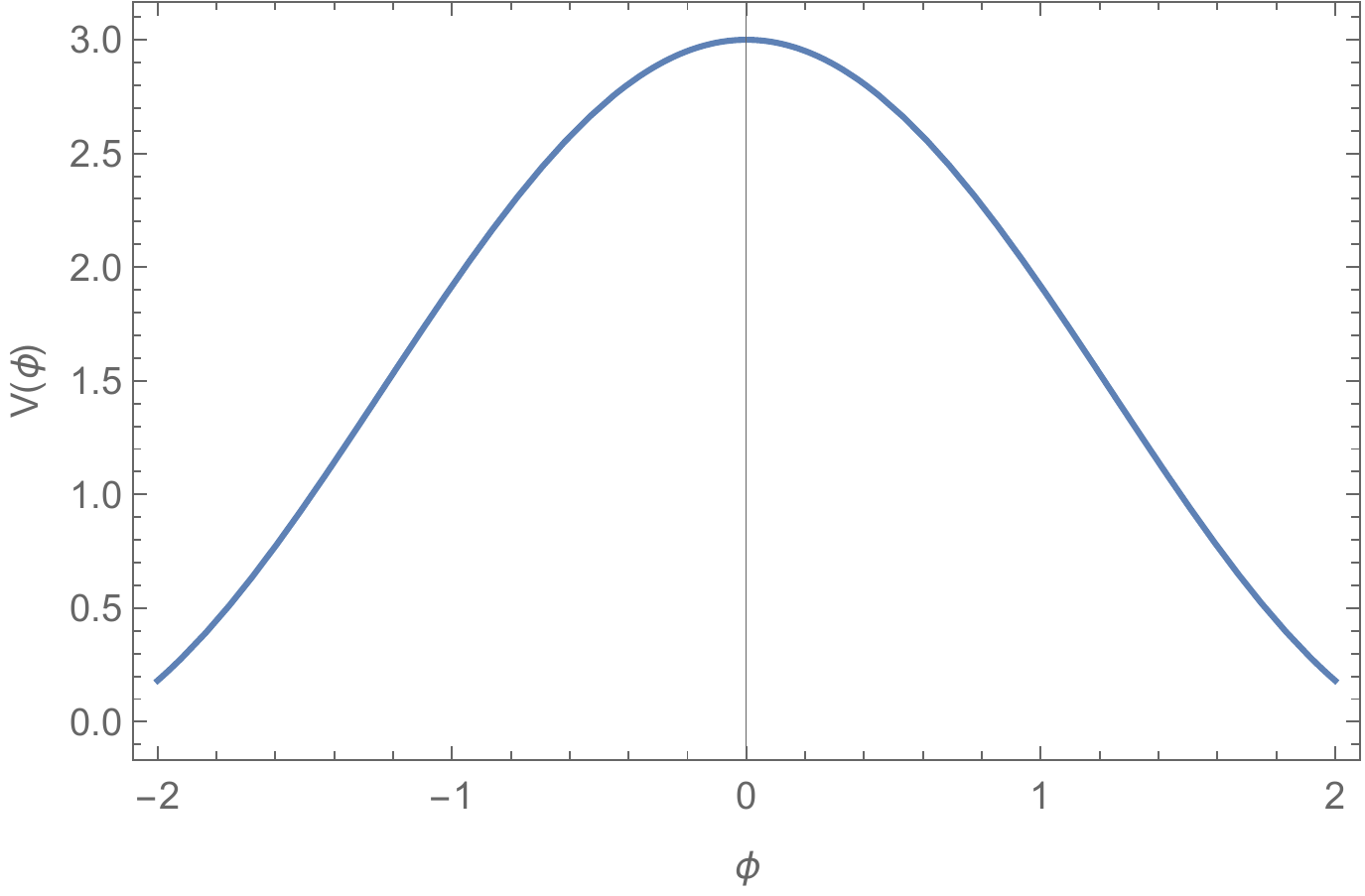}
	\caption{$H(t)$ vs $t$ plot with $\alpha=-3.5$, $m=14$, $\kappa=1$ and $C=1$.}
\end{minipage}
\end{figure}
The corresponding inflaton potential is given by, 
 \begin{eqnarray}
V(\phi)&=&\displaystyle \frac{3 C^2}{\kappa ^2} \left[1+\frac{\beta}{6}\left\lbrace 1-\cos(\sqrt{(m+1)|3+\beta|} \kappa \phi)\right\rbrace \right]\label{potential}
\end{eqnarray}
which becomes a constant i,e. $V(\phi)=\displaystyle \frac{3C^2}{\kappa^2}$ for $\beta=0$ and hence models of DBI inflation with constant potential. 
The plot of $V(\phi)$ vs $\phi$ depicted in fig. 12 shows that the inflaton potential is quadratic and inverted near the origin. 
Using (\ref{potential}) the critical value of the scalar field $\phi_{crit}$ can be determined such that $V(\phi=\phi_c)=0$. Then $\phi_{crit}$ is given by,
\begin{equation}
\phi_{crit}= \displaystyle \frac{1}{\kappa\sqrt{|3+\beta|(m+1)}}\cos^{-1}\left[1+ \frac{\beta}{6}\right]
\end{equation}
Now, for any $x$, $\cos^{-1} (x)$ is real and positive in the domain $0 \leq x \leq 1$. This indicates that for $\phi_{crit}$ to be real, the domain of validity of the DBI constant roll parameter is $-6\leq\beta\leq0 $. Thus the nature of the obtained potential resembles the hilltop inflation in the DBI theory where inflation occurs as the inflaton rolls down from the origin and ends as $V(\phi)$ vanishes for $\phi=\phi_c$. This form of potential is similar to the natural inflation model \cite{Freese}, however differs due to an additional constant.
The scalar field profile, Hubble parameter, scale factor and equation of state parameter are given by,
\begin{eqnarray}
\phi(t)&=& \frac{4}{\kappa\sqrt{(m+1)|3+\beta|}}\tan^{-1}(e^{\frac{(m+1)|3+\beta|}{2m}C t})  \label{cos-phi}\\[2mm]
H(t)&=&-C \tanh\left[\frac{(m+1)|3+\beta|}{2m} C t\right]\\[2mm]
a(t)&=&b_1 \cosh^{-\frac{(m+1)|3+\beta|}{2m}}\left[\frac{(m+1)|3+\beta|}{2m} C t\right] \\[2mm]
\omega(\phi)&=&-1+\frac{(m+1) \left| 3+\beta\right|  \tanh ^2\left(\frac{1}{2}  \sqrt{(m+1)\left| 3+\beta \right| } \kappa \phi\right)}{3
	m}
\end{eqnarray}
where $b_1$ is the constant of integration. In this case, the scalar field $\phi \rightarrow 0$ as $Ct \rightarrow -\infty$ whereas the at late time the inflaton field becomes almost constant as shown in fig. 13. The time evolution of the Hubble constant is shown in fig. 14.
\begin{figure}[h]
	\begin{minipage}{0.5\textwidth}
		\includegraphics[width=2.5 in]{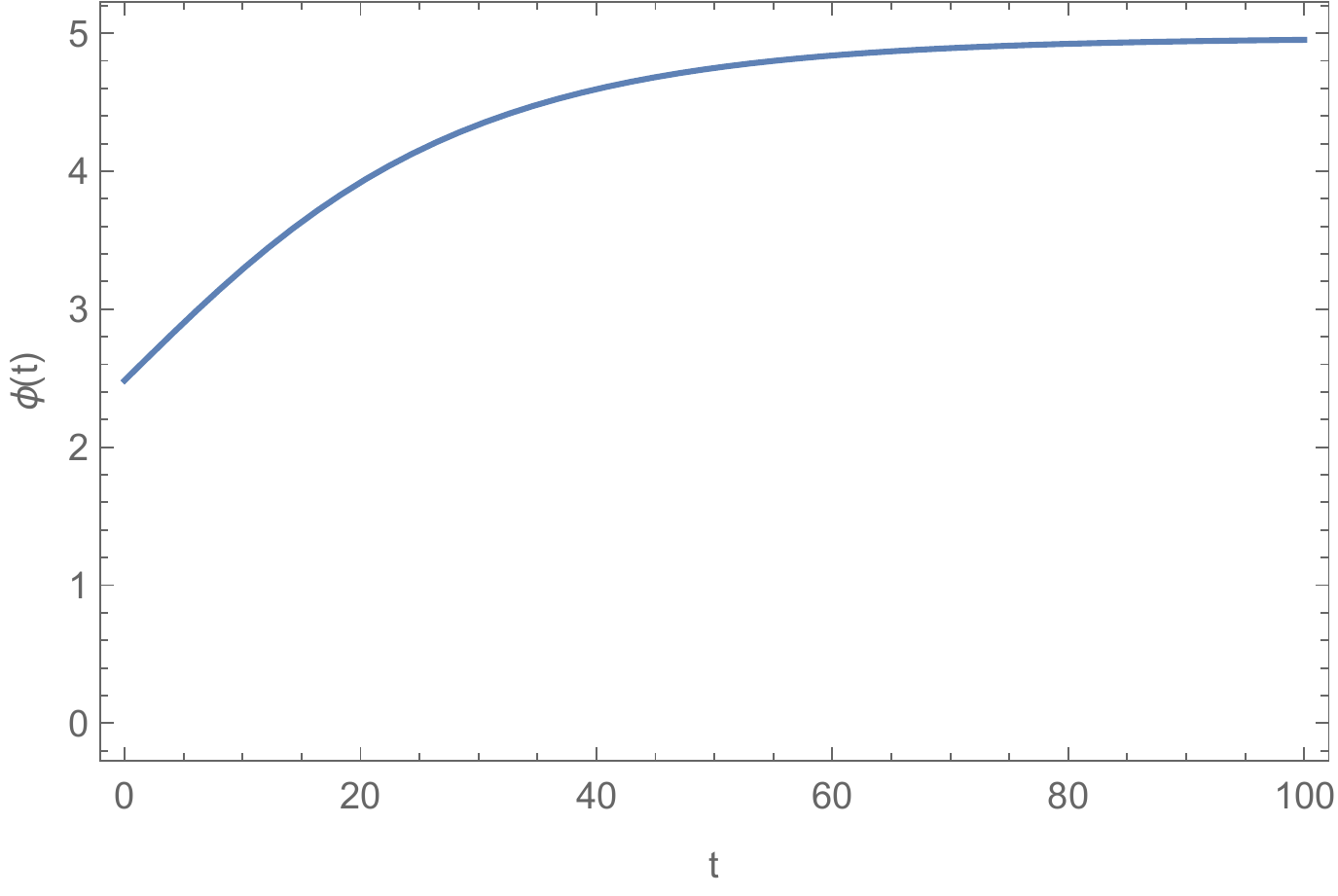}
		\caption{$\phi(t)$ vs $t$ plot with $m=14,\\ \beta=-3.5,\kappa=1$ and $C=1$.}		
	\end{minipage}
	\hfill
	\begin{minipage}{0.5 \textwidth}
		\includegraphics[width=2.5 in]{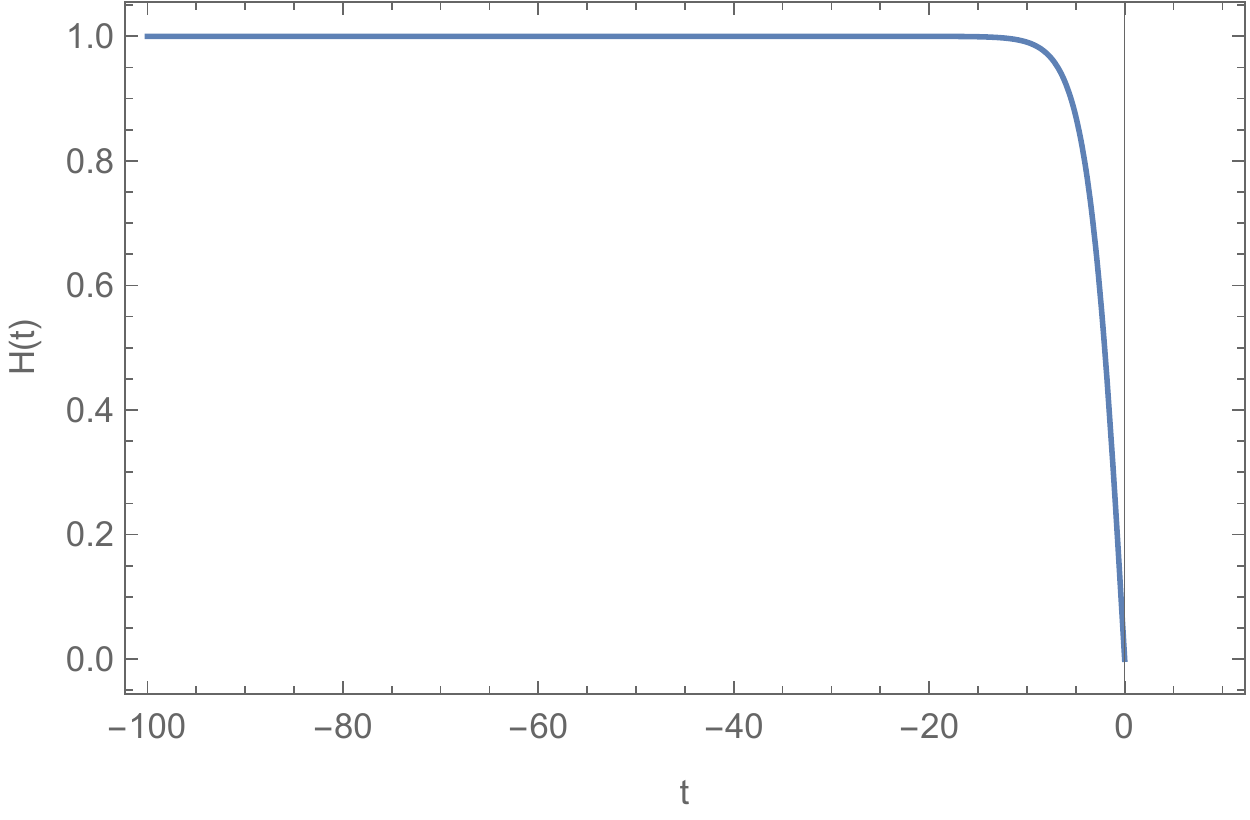}
		\caption{$H(t)$ vs $t$ plot with $\beta=-3.5$, $m=14$, $\kappa=1$ and $C=1$.}
	\end{minipage}
\end{figure}
In this case, the Hubble flow parameters $\epsilon_{H}$ and $\eta_{H}$ are given by,
\begin{equation}
\epsilon_{H}=\frac{(m+1) \left| 3+\beta \right| }{m \left[\cosh \left(\frac{ (m+1) \left|3+ \beta \right| }{m} C t\right)-1\right]}, \quad \eta_{H}=\frac{(m+1) \left|3+ \beta \right|  \coth ^2\left(\frac{(m+1) \left| 3+\beta \right| }{2 m} Ct\right)}{m}
\end{equation}
In the late time $Ct >>1$, $\epsilon_{H} \rightarrow 0$ whereas $\eta_{H} \sim {\cal O}(1)$ which indicates violation of slow roll approximations as shown in fig. 15 and fig. 16 respectively.
\begin{figure}[h]
	\begin{minipage}{0.5\textwidth}
		\includegraphics[width=2.5 in]{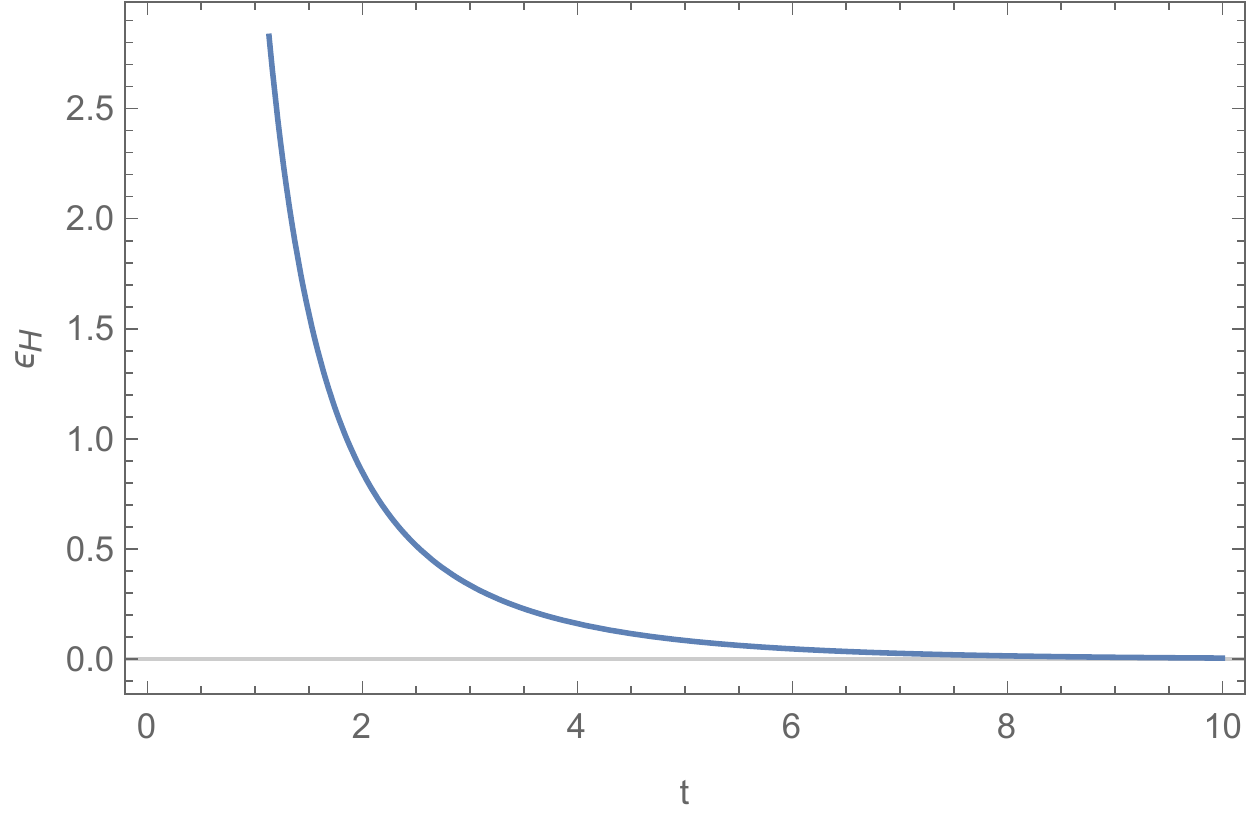}
		\caption{$\epsilon_{H}$ vs $t$ plot with $m=14,\\ \beta=-3.5,\kappa=1$ and $C=1$.}		
	\end{minipage}
	\hfill
	\begin{minipage}{0.5 \textwidth}
		\includegraphics[width=2.5 in]{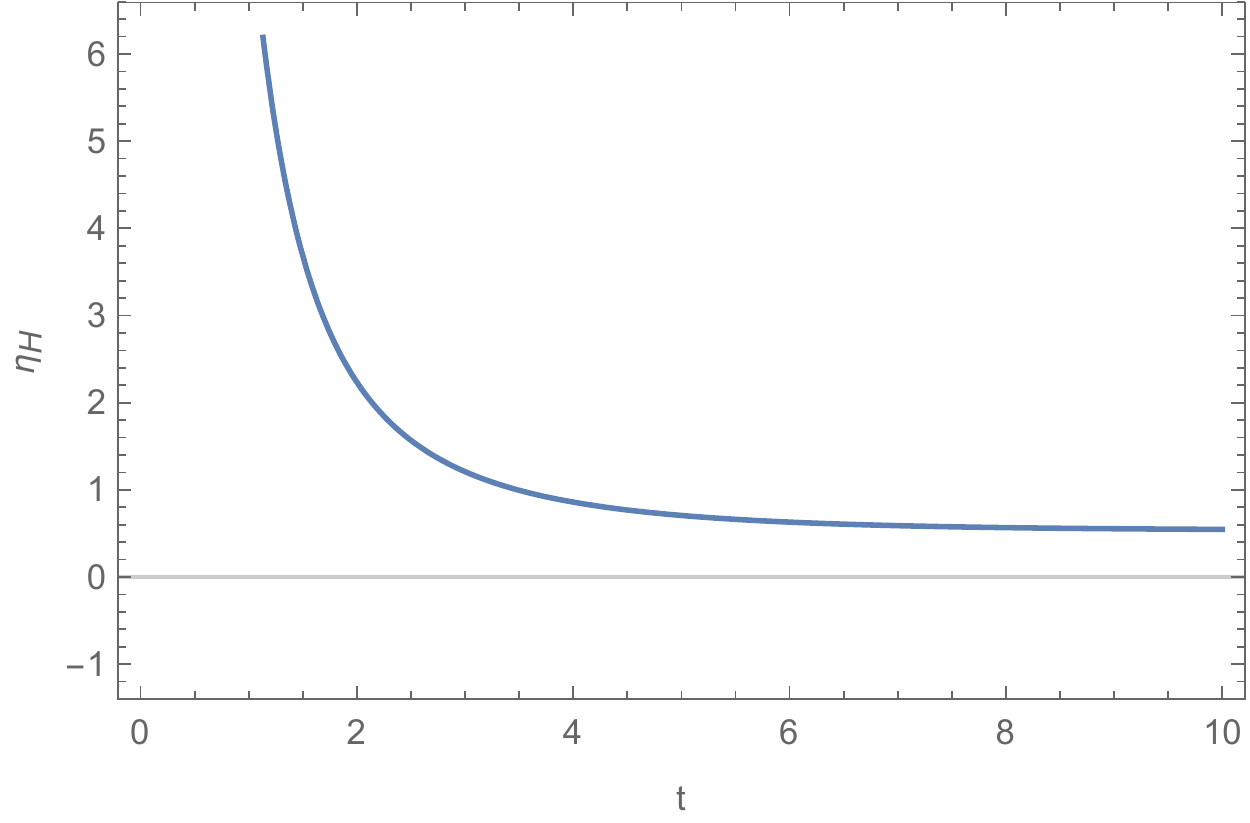}
		\caption{$\eta_{H}$ vs $t$ plot with $\beta=-3.5$, $m=14$, $\kappa=1$ and $C=1$.}
	\end{minipage}
\end{figure}
 Alternatively by computing the DBI flow parameters $\epsilon(\phi)$ and $\eta(\phi)$ using (\ref{Hubble3}),
 \begin{equation}
\epsilon(\phi)=\frac{(m+1) \left|3+ \beta \right|} {2m} \tanh ^2\left(\frac{1}{2} \sqrt{(m+1)\left|3+ \beta \right|  } \kappa \phi\right ), \quad \eta(\phi)=-\frac{(m+1) \left| 3+\beta \right| }{2 m}
\end{equation}
it can be readily seen in the small $\phi$ limit that although $\epsilon(\phi)$ is vanishingly small,   $\eta(\phi)$ is non-negligible constant during inflation. 
\subsection{Attractor solution}
 In the previous section, exact inflationary solutions are determined by employing the Hamilton-Jacobi method under constant roll conditions (\ref{6}) in the DBI scenario with constant $\gamma$. The solutions can be categorized into two different parameter spaces namely $\beta>-3$ and $\beta<-3$ where $\beta$ is the DBI constant roll parameter.
 We now explore the attractor behaviour of the obtained solutions by solving the equations of motion numerically with given inflaton potential for both of these cases.
In the $\beta>-3$ region, the cosine hyperbolic potential
$V(\phi)\rightarrow -\infty$ as $\phi \rightarrow \pm \infty$ which indicates the potential is unbounded from below if $\phi$ is very large. Such a solution is not an attractor.\\ 
Now with $\beta<-3$, the inflaton potential given by (\ref{potential}) remains finite and bounded for any as $\phi \rightarrow \pm \infty$. The $\phi$ vs $\dot{\phi}$ phase plot in this parameter space with $\gamma=m$ shows an attractor nature of the corresponding to potential (\ref{potential}) and warp factor (\ref{cos-warp}) for different initial conditions corroborated by the behaviour of each of the trajectories which converge to $\dot{\phi} \rightarrow 0$ and $\phi \rightarrow 0$ for $\beta=-3.03$ and $c_s=0.075$ which is taken in accordance to the limit imposed on the speed of the sound by Planck data (\cite{Planck1}).
\begin{figure}[h]
	\centering
	\includegraphics[width=6.0 in]{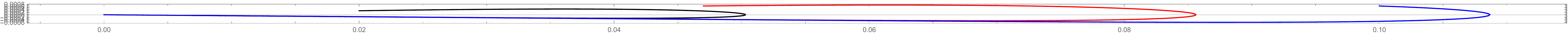}
	\caption{$\phi$-$\dot{\phi}$ plot with $\alpha=-3.03,\kappa=1$, $C=1$ and $\gamma^{-1}=c_s=0.075$.}
\end{figure}
\end{itemize}  
\subsection{Power spectra analysis}
In this section we study the power spectrum of linearized cosmological perturbations generated during inflation which is supported by constant roll conditions in presence of DBI term. 
Our analysis is based on the formulation by Garriga and Mukhanov \cite{Garriga} that deals with detailed study of generation of cosmological perturbations for general class of Lagrangian described by the non-canonical kinetic terms and permits variable speed of sound.
During inflation small field perturbations of the inflaton field in the quasi-de Sitter background are expressed up to linear order as,
\begin{equation}
\phi(t,x)= \phi_0(t)+\delta \phi (t,x)
\end{equation}
These inhomogeneities contribute to the perturbed FRW spatially flat metric which in the longitudinal gauge is given by,
\begin{equation}
ds^2=(1+2 \Phi)dt^2+(1-2 \Phi)a^2(t) \delta_{ij} dx^{i} dx^{j}
\end{equation} 
Here $\Phi$ is the Newtonian gravitational potential.  
The quantum fluctuations of the inflaton field  
as a consequence of accelerated expansion of the universe are stretched away outside the horizon radius where in general the field perturbations become non-dynamical. The field fluctuations are then manifested as classical metric perturbations which can be categorized into scalar (curvature) perturbations and tensor perturbations (gravitational waves). The curvature perturbations couple to matter which lead to structure formation of the universe and give rise to temperature fluctuations in cosmic microwave background radiation.  In the DBI scenario, the scalar fluctuations no longer travel with the speed of light, instead propagate at a reduced speed that of sound $c_s$ with an associated length scale $c_s H^{-1}$. 
We will concentrate our study on linearized primordial fluctuations of the inflaton field and in particular on scalar perturbations since the scalar fluctuations are only influenced by the presence of sound horizon whereas the tensor perturbations remain unaffected.
The gauge invariant curvature perturbation $\zeta$ related to the power spectrum is given by,
\begin{equation}
\zeta= \Phi+ \frac{H \delta \phi}{\dot{\phi}}
\end{equation}
such that $\delta \phi(t,x)$ and $\Phi$ are not independent. Defining a new variable $v=\zeta z $, 
the non-canonical generalization of the equation of motion of the quantum mode function $v_{k}$ for a comoving given wavenumber $k$ is given by Mukhanov equation \cite{Garriga} as follows,
\begin{equation}
\frac{d^2 v_{k}}{d \tau^2}+ \left(\frac{k^2}{\gamma^2}- \frac{1}{z^2}\frac{d^2 z}{d \tau^2}\right)v_{k}=0  \label{mode}
\end{equation} 
where $\tau$ is the conformal time related to the scale factor by the relation $d \tau= \displaystyle\frac{dt}{a}$.
The power spectrum of the curvature perturbations is given by,
\begin{equation}
P_{\zeta}(k)=\frac{k^3}{2 \pi^2} \displaystyle |\zeta_{k}|^2 = \frac{k^3}{2 \pi^2} \displaystyle |\frac{v_{k}}{z}|^2
\end{equation}
where the variable $z$ is defined as,
\begin{equation}
z= \frac{a \dot{\phi} \gamma ^{3/2}}{H} = -\frac{a \gamma \sqrt{2 \epsilon}}{\kappa}
\end{equation}
In the sub-Hubble region, 
the potential term $\displaystyle \frac{1}{z^2}\frac{d^2 z}{d \tau^2}$ can be neglected from (\ref{mode}) which then reduces to,
\begin{equation}
\displaystyle \frac{d^2 v_{k}}{d \tau^2}+ \frac{k^2}{\gamma^2} v_{k}=0
\end{equation}
and gives rise to plane wave solution as follows,
\begin{equation}
v_{k}=M_1 e^{-i k c_s \tau} + M_2 e^{i k c_s \tau}
\end{equation}
where $M_1$ and $M_2$ are integration constants. The fixing of two integration constants are carried out by choice of initial vacuum state of the fluctuations at the inflationary epoch and using normalization condition for the modes $v_k$. Adopting the standard approach of vacuum selection which corresponds to an adiabatic Minkowski vacuum in the past i,e. $t \rightarrow 0$/ $\tau \rightarrow -\infty$ with minimum energy, we take Bunch-Davis vacuum and impose $M_2=0$. The normalization condition corresponds to taking Wronskian condition for the mode as,
 \begin{equation}
 i\left[v_k^{*}\frac{d v_{k}}{d \tau}-v_k \frac{d v_{k}^{*}}{d \tau}\right]=1
 \end{equation} 
Then in the sub-Hubble region that corresponds to the the small wavelength limit, we obtain the mode solution as,
\begin{equation}
v_{k}=\frac{1}{\sqrt{2 c_s k }} e^{-i k c_s \tau}
\end{equation}
In order to determine the general solution of the mode equation for any comoving wavenumber k, let us express (\ref{mode}) in terms of flow parameters.
Then the potential term $\displaystyle \frac{1}{z^2}\frac{d^2 z}{d \tau^2}$ in terms of flow parameters as follows can be written as \cite{Tye},
\begin{eqnarray}
\frac{1}{z^2}\frac{d^2 z}{d \tau^2}=a^2 H^2 \left[(1+\epsilon-\eta-\frac{s_1}{2})(2-\eta-\frac{s_1}{2})+2 \epsilon(\epsilon-\eta-\frac{s_1}{2})\nonumber \right.\\
\left.
-\eta(\epsilon+s_1)+\xi-\frac{s_1}{2}(2s_1+\epsilon-\eta)+\epsilon s_2\right]  
\end{eqnarray}
Since $s_1=s_2=0$, the above equation reduces to,
\begin{eqnarray}
\frac{1}{z^2}\frac{d^2 z}{d \tau^2}=a^2 H^2 \left[2 +2 \epsilon- 3 \eta -4 \epsilon \eta + 2 \epsilon^2 + \eta^2 + \xi \right] \label{z-perturb}
\end{eqnarray} 
 We now define a variable $y$ as the ratio of the wavenumber $k$ to the sound horizon as,
\begin{equation}
y=\frac{k}{\gamma a H}= \frac{c_s k}{a H}
\end{equation}
Using (\ref{z-perturb}), the mode equation (\ref{mode}) becomes,
\begin{eqnarray}
(1-\epsilon)^2 y^2 \frac{d^2 v_{k}}{dy^2}+ (2 \epsilon^2-2 \epsilon  \eta)\frac{d v_{k}}{dy}+ \left[y^2-(2+2 \epsilon-3 \eta-4 \epsilon \eta+2 \epsilon^2+\eta^2+\xi)\right]v_{k}=0  \label{mod-eqn}\nonumber \\
\end{eqnarray}
The solution of this differential equation gives the solution of the quantum mode $v_{k}$ for any $y$.
The mode equation is solved by assuming that flow parameters evolve very slowly compared to the scale factor $a(t)$ so that the rate of change of these flow parameters with respect to the e-folding number is constant \cite{Kinney2}. Now from the exact relation,
\begin{equation}
dy=-k c_s(1-\epsilon)d \tau
\end{equation}
we can integrate over $d\tau$ to recast the quantum mode solution $v_{k}$ in the small wavelength as $v_{k}=\displaystyle \frac{1}{\sqrt{2 c_s k}}e^{\frac{iy}{(1-\epsilon)}}$ after imposing Bunch-Davis vacuum condition and fixing the normalization condition of the quantum modes from the Wronskian condition expressed in terms of $y$ as,
\begin{equation}
\left[v_k^{*}\frac{d v_{k}}{d y}-v_k \frac{d v_{k}^{*}}{d y}\right]=\frac{i}{kc_s(1-\epsilon)}   \label{bc-1}
\end{equation}
As the mode k approaches the Hubble radius crossing, $\displaystyle \frac{z''}{z}$ becomes important, therefore one has to solve the full equation (\ref{mod-eqn}) by taking into account of the potential term.
Since we are interested in obtaining curvature power spectra  generated during inflation at late times, the mode equation is to be solved by taking into account of the values of flow parameters $\epsilon_{H}$ and $\eta_H$ in the late time limit or equivalently with $\epsilon$ and $\eta$ computed in the small $\phi$ limit. At late times i,e. $C t>>1$,  $\epsilon_{H} \rightarrow 0$, $\xi_{H} \rightarrow 0$ whereas $\eta_{H}$  is non-negligible and remains almost constant in both the regions of $\beta<-3$ and $\beta>-3$. Alternatively, in both parameter regions, the DBI flow parameters show similar behaviour
\begin{eqnarray}
\epsilon \rightarrow 0, \qquad \eta= \frac{(m+1)}{2m}(3+\beta)   \label{9}
\end{eqnarray}
in the small $\phi$ limit. Then the mode equation (\ref{mod-eqn}) together with (\ref{9}) is expressed as,
\begin{equation}
 y^2 \frac{d^2 v_{k}}{dy^2}+\left[y^2-\left\lbrace 2+\frac{(m+1)(3+\beta)}{4m^2}\left(3(1-m)+\beta(m+1)\right)\right\rbrace \right]v_{k}=0
\end{equation}
whose solution is proportional to Hankel functions of the first and second kind as,
\begin{equation}
v_{k}(y)=\sqrt{y} \left[M_1 H_{\nu}^{(1)}(y)+M_2 H_{\nu}^{(2)}(y)\right]  \label{10}
\end{equation} 
We choose the Bunch-Davis vacuum to set $M_2=0$ and $M_1$ is fixed using the normalization condition (\ref{bc-1}) so that the solution of the mode function  for any given $y$ becomes,
\begin{equation}
v_{k}(y)=\sqrt{\frac{\pi}{4 k c_s}} \sqrt{y}H_{\nu}(y)
\end{equation}
where 
\begin{eqnarray}
\nu&=& \left| \frac{3+\beta(m+1)}{2m} \right|  \label{nu-1}
\end{eqnarray} 
which is a constant quantity.
Under different limits of $\gamma$, following models of inflation can be retrieved from the DBI constant roll inflation as,
\begin{itemize}
	\item Canonical constant roll inflation : $\gamma=m=1$, $s=0$, $\beta=\alpha$ and $\nu=\displaystyle \left| \alpha+\frac{3}{2} \right| $, where $\alpha$ is the canonical constant roll parameter.
	\item Canonical ultra-slow roll inflation : $\gamma=m=1$, $s=0$, $\beta=0$ and $\nu=\displaystyle \frac{3}{2} $.
	\item Ultra slow-roll DBI inflation with constant speed of sound : $\gamma=m \neq 1$, $s=0$, $\beta=0$ and $\nu=\displaystyle \frac{3}{2m} $.
\end{itemize}  
 The power spectrum of curvature perturbations can be determined in the large wavelength limit which corresponds to $y \rightarrow 0$. Then,
\begin{eqnarray}
P_{k}&=& \frac{k^3}{2 \pi^2}|\frac{v_{k}}{z}|^2_{y\rightarrow 0} \nonumber  \\
&=&\frac{k^3}{2 \pi^2} \frac{\pi}{4 k c_s} y \frac{H^2}{a^2 \dot{\phi}^2 \gamma^3} | \lim_{y \to 0} H_{\nu}(y) |^2 \nonumber
\end{eqnarray}
As $ \lim_{y \to 0} H_{\nu}(y) \simeq \displaystyle \frac{-i}{\pi} \Gamma(\nu) \left(\frac{y}{2}\right)^{-\nu}$, the power spectra of curvature perturbations is found to be, 
\begin{eqnarray}
P_{k}&=& \frac{H^4}{4 \pi^2 \dot{\phi}^2}2^{2\nu-3} \frac{\Gamma(\nu)}{\Gamma(3/2)}y^{3-2\nu}
\end{eqnarray} 
The scalar spectral index of curvature perturbations is then by,
\begin{eqnarray}
n_s-1&=&3-2 \nu \nonumber \\
&=&3- \left|  \frac{3+\beta(m+1)}{m} \right|  \label{nu}
\end{eqnarray}
We now estimate the DBI constant roll parameter using $n_s$ and speed of sound $c_s$. From Planck and WMAP results \cite{Planck1}, the speed of sound for DBI models is constrained in the range $0.07<c_s<1$ and the scalar spectral index lies within the interval $0.93\leq n_s\leq 0.96$. Assuming the v $n_s = 0.96$ and using (\ref{nu}) we get,
\begin{equation}
	\beta= -\frac{m(0.96-4)+3}{(m+1)}, \qquad \beta= -\frac{m(0.96-4)-3}{(m+1)}
	\end{equation}
Taking $c_s = 0.075$ we obtain $\beta=2.61$ and  $\beta=-3.037$ respectively. But $\beta =2.61$ lies in the region $\beta>-3$ and does not give rise to attractor solution whereas it has been already shown that inflationary solution corresponding to $\beta<-3$ is an attractor hence $\beta=2.61$ is discarded. This shows that the DBI constant roll parameter $\beta$ in the constant $\gamma$ case remains close to its canonical counterpart \cite{Starobinsky1}. As $c_s$ approaches unity starting out from $c_s=0.075$, $\beta$ initially taking the contributions of DBI term slowly approaches its canonical limit which can be seen from the figure 18.
\begin{center}
\begin{figure}
	\centering
		\includegraphics[width=2.5 in]{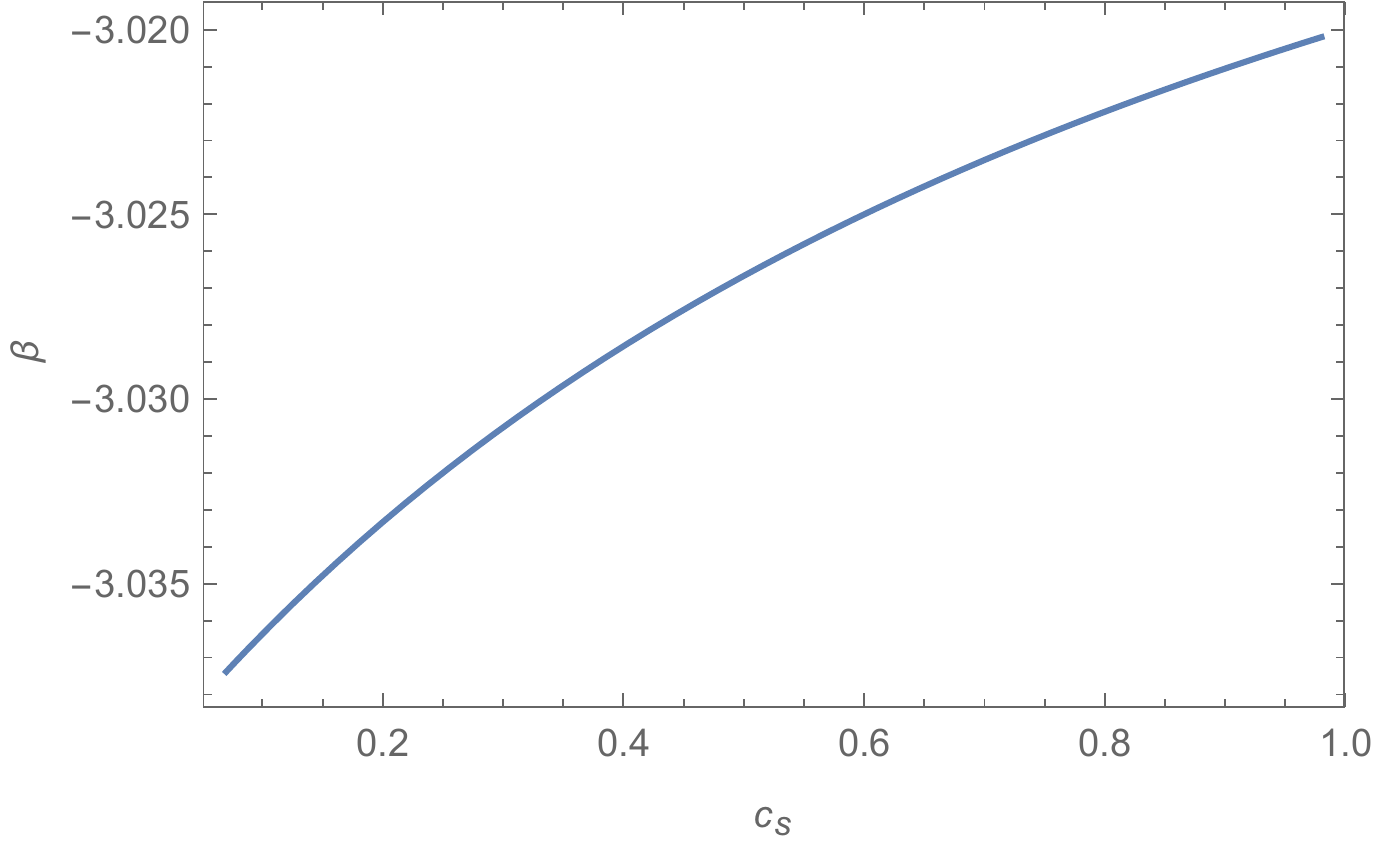}
		\caption{Plot of $c_s$ vs $\beta$ for $n_s =0.96$. $\beta$ approaches the canonical value $-3.02$ as $c_s \rightarrow 1$.}
	\end{figure}
\end{center}
\section{Discussions}
In the present paper, we have generalized the constant roll inflation model in the non-canonical framework without employing slow-roll conditions. For this, we have assumed scalar DBI theory which in phenomenological sense may be included in k-inflation models involving non-standard kinetic terms. 
Inflationary scenarios addressed using Dirac-Born-Infeld model at the same time provides a framework for string theory realization of inflation due to relativistic motion of a D3 brane in warped geometry subject to constant roll conditions. \\
In the DBI regime, we have determined the condition for sustaining constant-roll inflation in presence of a time-independent speed of sound. However, for determining analytical solutions, we assumed the simplest case i,e. the constant speed of sound of inflaton fluctuations and obtained inflationary solutions which are categorized in two parameter regions associated with the DBI constant roll parameter $\beta$.
In the given scenario subject to constant roll conditions we obtained DBI power-law inflation, variant of natural inflation and cosine hyperbolic solution. Corresponding nature of DBI warp factor characterising the throat geometry, the time evolution of the scalar field profiles and nature of evolution of scale factors for each of the obtained inflationary solution have been studied. However, among all inflationary solutions, the natural inflation turns out to be an attractor solution.
We also studied the curvature power spectra for constant-roll DBI inflation and determined the scalar spectral index. Finally we estimated the allowed value of the DBI constant roll parameter using present observational bounds on speed of sound $c_s$ and spectral index $n_s$. 

\section*{Acknowledgements}
I would like to thank Narayan Banerjee for useful discussions. I am thankful to Jiro Soda, Dominik Schwarz and Volker Perlick for suggestions in the earlier stage of this work. This work is supported by German Research Foundation DFG grant/40400957.

\end{document}